\documentclass[11pt,a4paper]{article}
\usepackage{amssymb}
\usepackage{amsmath}
\usepackage{jheppub}

\newcommand{\be}{\begin{equation}}
\newcommand{\ee}{\end{equation}}
\newcommand{\bs}{\begin{split}}  
\newcommand{\es}{\end{split}}     
\newcommand{\beqa}{\begin{eqnarray}}
\newcommand{\eeqa}{\end{eqnarray}}

\newcommand{\mP}{\mathcal{P}}
\newcommand{\mH}{\mathcal{H}}
\newcommand{\dd}{\text{d}}
\newcommand{\vE}{\mathbf E}
\newcommand{\vB}{\mathbf B}
\newcommand{\cH}{\mathcal{H}}
\newcommand{\cP}{\mathcal{P}}
\newcommand{\vk}{\mathbf{k}}
\newcommand{\vp}{\mathbf{p}}
\newcommand{\vq}{\mathbf{q}}

\title{Statistics of Anisotropies in Inflation with Spectator Vector Fields 
}

\author[a]{Mikjel Thorsrud,}
\author[b]{Federico R.\ Urban,}
\author[a]{David F.\ Mota}

\affiliation[a]{Institute of Theoretical Astrophysics, University of Oslo, \\
P.O. Box 1029 Blindern, N-0315 Oslo, Norway.}
\affiliation[b]{Service de Physique Th\'eorique, Universit\'e Libre de Bruxelles, \\
CP225, Boulevard du Triomphe, B-1050 Brussels, Belgium.}
\emailAdd{mikjel.thorsrud@astro.uio.no}
\emailAdd{furban@ulb.ac.be}
\emailAdd{d.f.mota@astro.uio.no}

\abstract{We study the statistics of the primordial power spectrum in models where massless gauge vectors are coupled to the inflaton, paying special attention to observational implications of having fundamental or effective horizons embedded in a bath of infrared fluctuations.  As quantum infrared modes cross the horizon, they classicalize and build a background vector field.  We find that the vector experiences a statistical precession phenomenon. Implications for primordial correlators and the interpretation thereof are considered.  Firstly, we show how in general two, not only one, additional observables, a quadrupole amplitude and an intrinsic shape parameter, are necessary to fully describe the correction to the curvature power spectrum, and develop a unique parametrization for them.  Secondly, we show that the observed anisotropic amplitude and the associated preferred direction depend on the volume of the patch being probed.  We calculate non-zero priors for the expected deviations between detections based on microwave background data (which probes the entire Hubble patch) and large scale structure (which only probes a fraction of it).}

\begin{document}

\maketitle

\section{Introduction}

One of the most intriguing observational consequences inflation with spectator vector fields is that it quite generally predicts anisotropic corrections to primordial correlation functions such as the curvature power spectrum (see, e.g., \cite{Ackerman:2007,Yokoyama:2008xw, Karciauskas:2008bc, Himmetoglu:2009mk, Watanabe:2010fh, Dulaney:2010sq, Gumrukcuoglu:2010yc, Karciauskas:2011fp, Emami:2011yi, Lyth:2012vn, Yamamoto:multivector2, Lyth:2012br, Soda:2012zm, Maleknejad12, Abolhasani13, Lyth:2013sha} and references therein). Searches for imprints in cosmological data have largely been plagued by systematics and/or foregrounds, leaving the empirical situation unclear \cite{Groeneboom:08,Hanson:09,Groeneboom:09,Pullen:2010}. Very recently, however, a null-detection with two-percent order error bars has been reported based on foreground-cleaned and beam-corrected sky maps from the Planck satellite \cite{komatsu} (see also \cite{Ramazanov:2013wea} for the most updated WMAP9 constraint). This leaves the status of a primordial anisotropy more or less similar to the related issue of non-Gaussanities: consistent with null and with decreasing error bars. Still, in the era of high-precision cosmology theoretical and observational effort will continue to constrain the space of theories in order to learn which type of interactions in the early universe are phenomenologically viable.

In models with vector fields coupled directly to the inflaton, the amplitude of the primordial anisotropy is typically determined by the fundamental parameters describing the interaction Lagrangian, which are therefore precious to explore the viability of a particular high-energy theory.  The power spectrum in this setup is normally parametrized by an axisymmetric quadrupole correction to the isotropic monopole term. Such a quadrupole provides three new empirical parameters: an amplitude $g(k)$ and a preferred direction $\hat{\mathbf n}$ (described for instance by two spherical coordinates $\vartheta$ and $\varphi$) defining the axis of rotational invariance. However, in a universe which at late times is, or looks, isotropic at the background level, there is no way to predict the preferred direction of perturbations. Consequently, the only additional parameter connecting theory to observations is the amplitude $g(k)$.  The axial symmetry of the quadrupole is a consequence of the symmetries of the vector field that is present during inflation. Thus, although the quadrupole correction is anisotropic, there remains one rotational symmetry whose axis is set by the direction of the vector. Nevertheless, if the vector field rotates / precesses as inflation proceeds, or if randomly oriented multiple vector fields are present, it is natural to expect the residual symmetry to be broken. 

In this paper we shall consider vector models coupled to the inflaton via the Lagrangian $I^2(\phi) F^2$, where $I(\phi)$ is a scalar function of the inflaton field $\phi$, and $F$ is the strength of a (single) gauge vector $A_\mu$. 
If there is only one vector field, there exists an anisotropic Bianchi type I attractor with a stable non-rotating background vector field $\vE_{0}$ \cite{attractor1,attractor2,attractor3,attractor4,Yamamoto:multivector1,attractor5,attractor6}. However, quite recently it was pointed out that the aforementioned attractor solutions are unstable with respect to quantum corrections \cite{peloso12}.  Quantum fluctuations of the vector field will continuously be transferred from the ultraviolet to superhorizon scales where they become classical (commuting) fields. Therefore, in addition to the electric-type vector $\vE_{0}$, which is the solution of the classical equations, there will be a contribution $\vE_{IR}$ from super horizon fluctuations which modulates the primordial power spectrum via loop terms.\footnote{We refer to ``electric'' and ``magnetic'' types in analogy with Maxwell theory, but we do not restrict our gauge fields to be the the Standard Model $U(1)$.}  It was shown in \cite{peloso12} that the the effective interaction Hamiltonian associated with $\vE_{IR}$ is structurally similar to the one of $\vE_0$, and that (in the single vector setup) it leads to significant corrections to primordial correlators.  In \cite{Thorsrud:2013mma} we explored observational consequences in the multivector context with a background attractor solution free of anisotropic stress, and showed that the isotropy of the underlying model is spontaneously broken by long wavelength gauge modes.  In this setup structurally new types of signatures appear and we showed that there exists a new local observable, $\chi$, which is exactly scale invariant and measures the quadrupole's departure from axial symmetry. For a general-type quadrupole, lacking any rotational symmetry, this parameter goes hand in hand with the amplitude $g(k)$ and both must be considered to fully link a theoretical model to observations at the two-point level. The first purpose of this paper is to provide a complete discussion of this parametrization, first employed in \cite{Thorsrud:2013mma}, including a proof of its mathematical consistency.

Next, we want to further discuss and identify observational implications of the IR vector. Although $\vE_{IR}$ is quasi-homogeneous, in the sense that its fluctuations cannot be probed locally within a Hubble patch, it varies over length scales much larger than the causal horizon. This leads to a landscape picture where primordial correlators depend on the local realization of the infrared vector or, equivalently, the position of the observer within the entire inflated space. Predictions for observables thence become a statistical problem not only depending on the parameters of the high energy theory, but also on those describing the observer, or more precisely, the data sets accessible to the observer. 
Moreover, $\vE_{IR}$ does precess since it is progressively built up as a Gaussian random walk by modes crossing the horizon. One of the goals of this paper is to understand the observational implications of this phenomenon. As mentioned above, na\"ively one would expect the axial symmetry of the quadrupole in single vector models to be broken by this rotation.  However, as we shall see, to good accuracy the symmetry is protected by an effective cut-off excluding modes whose wavelength is smaller than the patch which the cosmic data is collected from.  Instead, the primary observational consequence is that different cosmological data sets, whose redshift coverage differ, such as the cosmic microwave background (CMB) versus large scale structure (LSS), will be expected to bear statistically different signatures.  This is an interesting observational consequence of having fundamental or effective horizons embedded in a bath of infrared fluctuations. We aim at studying the effect at the quantitative level, both for single vector models and multi-vector models.

To summarize, the main goal of this paper is to further advance our understanding of anisotropic signatures in the context of models with Lagrangians of the form $I^2(\phi)F^2$.  In this framework we will discuss in depth the statistical properties of the anisotropies characterizing the quadrupolar correction, and connect, again statistically, theoretical expectations with measurements in different datasets / probes.  We shall start with a full description of our (fairly model-independent) parameterization of the primordial power spectrum in section \ref{chparameterizations}.  Then, in section \ref{chback}, we introduce the background model (section \ref{chmodel}), and summarize known results about vector quantum fluctuations (section \ref{chvec}) and their power spectra (section \ref{chpower}).  With the tools at hand we will discuss the lanscape picture which derives from our setup in section \ref{chbubble}; we include a detailed physical interpretation of the spectrum and the role of the IR fluctuations (section \ref{chircut}) including phenomenological constraints  (section \ref{chplanck}) and generalizations to the multi-vector case and the three-point function (section \ref{3pf}).  Section \ref{chstatistics} unravels the statistical properties of the precession phenomenon (section \ref{chstats}), which we concretely apply (section \ref{chcmblss}) to give priors for deviation in detections based on different data sets in both single and multivector models.  Finally, in section \ref{chconc} we summarize our main findings.

\section{Parametrizing the primordial power spectrum \label{chparameterizations}}

In this section we shall develop a parametrization of general quadrupolar power spectrums which isolates the intrinsic degrees of freedom we can make predictions for.  The power spectrum $\mathcal{P}_\zeta(\mathbf k)$ is defined via the two-point correlation in Fourier space:
\be
\left< \zeta(\mathbf k)\; \zeta(\mathbf p)  \right> = 2\pi^2 \frac{\delta^3(\mathbf k + \mathbf p)}{k^3}  \mathcal{P}_\zeta(\mathbf k),
\ee
where the Dirac-delta function is a consequence of statistical translational invariance.\footnote{The curvature perturbation $\zeta$ we employ is the comoving curvature perturbation evaluated in flat gauge (there the spatial scalar perturbation is set to zero); on superhorizon scales it nearly corresponds to the uniform density curvature perturbation, so that our results translate directly from one to the other.}  Since we do not assume rotational invariance, the power spectrum $\mathcal{P}_\zeta (\mathbf k)$ depends not only on the magnitude $k=|\mathbf k|$ of the wave vector, but also on its direction \newline $\hat{\vk}=(\sin\vartheta \cos\varphi, \sin\vartheta \sin\varphi, \cos\vartheta)$. We parametrize the most general statistically homogenous power spectrum in the following form
\be 
\mathcal{P}_\zeta (\mathbf k) =  \mathcal{P}_\mathcal{I}(k) \left( 1 + \sum_{l=\text{even}} \sum_{m=-l}^{l} b_{lm}(k) Y_{lm} (\vartheta,\varphi) \right),
\label{formidable}
\ee
where $Y_{lm}(\vartheta,\varphi)$ are the spherical harmonics. Note that $\mathcal{P}_\mathcal{I}(k)$ is the isotropically averaged power spectrum:
\be
\int_0^{2\pi} \int_0^{\pi} \mathcal{P}_\zeta (\mathbf k) \sin\vartheta d\vartheta d\varphi = \mathcal{P}_\mathcal{I}(k). 
\ee
In this paper we use real $Y_{lm}$'s so that all coefficients $b_{lm}$ are real. The monopole term ($l=0$) is accounted for by the first term in the parenthesis, while ($l=2,4\dots$) account for the quadrupole and higher order terms. The odd terms $(l=1,3\dots)$ vanish because of the reality condition $\zeta(-\mathbf k)=\zeta^*(\mathbf k)$ and hence $\mathcal P_\zeta(\mathbf k)=\mathcal P_\zeta(-\mathbf k)$.  In general the coefficients $b_{lm}$ are functions of $k$ which means that the anisotropy is scale dependent.

In models of inflation with spatial vectors, the power spectrum is commonly parametrized in the axisymmetric form:
\be
\mathcal P_\zeta (\mathbf k) = \mathcal P_\mathcal{I}(k) (1+g(k) \cos^2 \vartheta_{\hat{\vk}, \hat{\mathbf n}}),
\label{planarsymmetric}
\ee  
where $\vartheta_{\mathbf k, \mathbf n}$ is the angle between the mode vector $\mathbf k$ and a preferred direction $\hat{\mathbf n}$
\be
\cos \vartheta_{\hat{\vk}, \hat{\mathbf n}} = \hat{\vk} \cdot \hat{\mathbf n}
\ee
and $g$ (in general a function of $k$) parametrizes the anisotropy. Although this power spectrum breaks full rotational invariance, it has one residual symmetry with respect to the axis $\hat{\mathbf n}$, which is assumed to be aligned with the vector field.   If the vector field rotates, or if multiple vector fields are present, the residual rotational symmetry will in general be broken.  The most general quadrupole is parametrized by five independent functions $b_{2m}(k)$. We shall now discuss an important subclass of this.

In the important special case when all five coefficients are proportional $b_{2i}(k)\propto b_{2j}(k)$, $\forall i \neq j \in \{-2,-1,0,1,2\}$ the power spectrum 
\be
\mathcal{P}_\zeta (\mathbf k) = \mathcal P_\mathcal{I}(k) \left( 1 + \sum_{m=-2}^{2} b_{2m} Y_{2m} (\vartheta,\varphi) \right),
\label{Qnotrotated}
\ee
can be simplified by taking into account that three of them correspond to the orientation of the quadrupole on the sky. By choosing an appropriate coordinate system, we can therefore set three of the five $b_{2m}$ coefficients to zero. More specifically, given an arbitrary state vector $(b_{2m})$ one can identify three Euler angles ($\psi_1,\psi_2,\psi_3$) and perform a rotation $(\tilde b_{2m}) = [R(\psi_1,\psi_2,\psi_3)] (b_{2m})$ so that $\tilde b_{2m}(k)=0$ for three of the five $m$'s.   In this case the remaining two coefficients represent the intrinsic shape of the quadrupole.  A two-dimensional parametrization is possible if we set the coefficients with $m=\{-2,-1,2\}$ to zero, so that the power spectrum can be written as a linear combination of $Y_{20}$ and $Y_{21}$. The Euler angles that set $\tilde b_{2-2}=\tilde b_{2-1}=\tilde b_{22}=0$ are not unique, but leave freedom to choose the sign of $\tilde b_{21}$ whereas $\tilde b_{20}$ and $|\tilde b_{21}|$ are unique. To remove the ambiguity we choose $\text{sign}(\tilde b_{21})=-\text{sign}(\tilde b_{20})$.  With this convention the mapping from five arbitrary coefficients $\{ b_{2m} \}$ to $\{\tilde b_{20},\tilde b_{21}\}$ is unique: for each set $\{ b_{2m} \}$ there is one and only one doublet $\{\tilde b_{20},\tilde b_{21}\}$. The proof is given in appendix \ref{chproof}.  We can therefore parametrize the power spectrum (\ref{Qnotrotated}) as
\be
\cP(\vk) = \cP_0(k)\!\left[1 + g(k)\! \left( \cos{\chi} A(\hat{\mathbf{k}}) + \sin{\chi} B(\hat{\mathbf{k}}) \right)\right],
\label{scaleinvariant}
\ee
where the directional functions 
\begin{align}
A(\hat{\mathbf{k}})&=\cos^2\vartheta-1/3, \\ 
B(\hat{\mathbf{k}})&=\sin2\vartheta \cos\varphi / \sqrt3, 
\end{align}
are proportional to $Y_{20}(\hat{\vk})$ and $Y_{21}(\hat{\vk})$, respectively. The parameters $g(k)\in \mathbb{R} $ and $\chi \in [0,\pi/2]$ are defined by
\begin{align}
g(k) &= \text{sign}(\tilde b_{20}) \frac{3}{4}\sqrt{\frac{5}{\pi}} \sqrt{|\tilde b_{20}|^2+|\tilde b_{21}|^2}, \label{gdef} \\
\cos{\chi} &= \frac{|\tilde b_{20}|}{\sqrt{|\tilde b_{20}|^2+|\tilde b_{21}|^2}}, \label{chi1def} \\
\sin{\chi} &= \frac{|\tilde b_{21}|}{\sqrt{|\tilde b_{20}|^2+|\tilde b_{21}|^2}}. \label{chi2def}
\end{align}

Hence, we have developed a generalization of the axially symmetric power spectrum (\ref{planarsymmetric}) which is not restricted by any continuous rotational symmetry.  The parametrization is valid under the assumption that all of the spherical harmonic coefficients have the same scale dependence.\footnote{If this is not the case, it is not possible to remove any of the coefficients by a global rotation since it would only be possible to set three of them to zero for a specific value $k=k_*$. When they are all proportional, however, an arbitrary quadrupole can be parametrized as (\ref{scaleinvariant}).} This condition is commonly obeyed by models of anisotropic inflation including interactions of the type $I^2(\phi)F^2$ which we focus on in this paper. Note that $g(k)$ has the same scale dependence as the functions $b_{2m}(k)$ while $\chi$ is exactly scale invariant. Also note that $\mathcal P_\zeta (\mathbf k)$ is invariant under a parity flip $\mathbf k \rightarrow -\mathbf k$, or equivalently $(k,\vartheta,\varphi)\rightarrow (k,\pi - \vartheta,\varphi+\pi)$, as required by the reality condition $\zeta(-\mathbf k)=\zeta^*(\mathbf k)$.  The absolute value of $g$ is an invariant under rotations and related to the quadrupole scalar by
\be
\mathcal{Q} \equiv  \left(\sqrt{\frac{1}{2\pi} \frac{l(l+1)}{(2l+1)}  \displaystyle\sum_{m=-l}^{l}|b_{lm}|^2} \right)_{l=2} = \frac{4}{5\sqrt{3}} |g|.
\ee 

For the case $\chi=0$ equation (\ref{scaleinvariant}) can be rewritten on the axisymmetric form (\ref{planarsymmetric}) with the redefinition $g/(1-g/3) \rightarrow g$. Planck data constrains $|g|$ to be maximum at the $2 \%$ level in which case the two definitions match well \cite{komatsu}.  Note, however, that there is no data analysis considering the two parameters family we have introduced here, thus the bound's dependency on $\chi$ is unknown. Nevertheless, it is reasonable to assume that the amplitude $|g|$ must be small also for $\chi \neq 0$.  The contribution from $Y_{20}$ and $Y_{21}$ are equivalent when $\chi=\pi/4$.  For $\chi\ll\pi/4$ the quadrupole is almost axisymmetric while for $\chi>\pi/4$ it is dominated by the symmetry-breaking term. 

Finally, it is important to point out that although there are just two parameters in our scheme, an observer looking for imprints in the CMB or LSS will have three additional parameters associated with the orientation of the quadrupole on the sky. It is in general not possible to make predictions for these additional parameters since they are drawn from uniform probability distributions. They merely correspond to the three Euler angles used to cast the power spectrum in the form (\ref{scaleinvariant}), where we have isolated the two intrinsic degrees of freedom that we can make predictions for, namely the amplitude $g$ and the shape $\chi$. Nevertheless, a data scrutator carrying out a likelihood analysis needs to trace out a five-dimensional parameter space. This is two more than if restricting to axisymmetric power spectrums on the form (\ref{planarsymmetric}).  In that case there is just one intrinsic parameter, namely the amplitude $g$, which is the only one that concrete models of inflation can forecast.  The two other empirical parameters correspond to the random orientation of the preferred direction on the sky: since the quadrupole is axisymmetric in this case, only two parameters are needed to fix it, adding up to an overall count of three.  We note that by breaking the axial symmetry of the quadrupole, we actually introduced two more parameters, one intrinsic ($\chi$) and one additional Euler angle required fix the orientation on the sky unambiguously.

\section{Vector fluctuations and the primordial power spectrum \label{chback}}

We summarize here all the basic definitions, conventions, and known results for our setup.

\subsection{Lagrangian and background dynamics \label{chmodel}}
The basic interaction Lagrangian we employ is
\be\label{lag}
\mathcal{L}_\text{int} = -\frac{1}{4} I^2(\phi) F_{\mu\nu} F^{\mu\nu},
\ee
where $F_{\mu\nu} = \nabla_\mu A_\nu - \nabla_\nu A_\mu$ is the strength of the vector field $A_\mu$.  The coupling is dynamical and depends explicitly on the inflaton field $\phi$ through $I(\phi)$.  We define the electric-type and magnetic type three vectors
\be
\vE = -\frac{\left<I\right>}{a^2}  \mathbf A', \qquad  \vB = -\frac{\left<I\right>}{a^2} \nabla \times \mathbf A,
\label{decompositionE}
\ee
where primes denote differentiation with respect to conformal time $\tau$, and we rely on the, isotropic, Friedmann-Lema\^itre-Robertson-Walker (FLRW) metric
\be\label{metric}
\dd s^2 = a^2(\tau) \left[- \dd\tau^2 + \dd\mathbf{x}^2 \right] \, .
\ee
Throughout this paper $\epsilon$ is the slow-roll parameter defined via $\epsilon {\cal H}^2 \equiv {\cal H}^2 - {\cal H}'$, where ${\cal H} \equiv a'/a \equiv aH$ is the conformal (comoving) Hubble parameter. $\mathbf A$ denotes the physical (transverse) three vector in the gauge $A_0=0$. $\mathbf E$ and $\mathbf B$ represent the (comoving) orthonormal frame components of the field strength tensor (times the modulation function) so that the energy density associated with the vector takes the standard form $(\mathbf E^2+\mathbf B^2)/2$. As mentioned in the introduction, we do not assume the gauge field to coincide with standard model $U(1)$ photons, but keep the electromagnetic notation for convenience.

In principle one should adopt a metric which can accomodate the anisotropy (of a homogenous vector) and study the perturbations in that background.  Although this is certainly not a prohibitive task, it does result in a fair computational complication since in that case scalar, vector, and tensor perturbation do mix.  However, as it was shown in \cite{Dulaney:2010sq,Gumrukcuoglu:2010yc,Watanabe:2010fh}, a percent level quadrupole anisotropy amplitude is generated by a tiny anisotropic expansion with $\Delta H / H \sim 10^{-8}$.  Moreover, the statistical anisotropy of perturbations comes essentially from the direct coupling (\ref{lag}) and not from exciting metric perturbations. Hence a Bianchi type background metric would not change the physical results of this work neither qualitatively nor quantitatively.   This prompts us to keep the isotropic metric for our work, following \cite{peloso12} (where an extensive discussion about this point can be found).

In this work we shall not rely on any specific form of the modulation function $I(\phi)$.  Instead we shall assume that it produces a scale invariant spectrum of frozen electric type vector modes. This is obtained when the vacuum expectation value scales as $\left<I\right>\propto a^{-2}$, in which case the magnetic type modes quickly decays (similarly, constant magnetic-type modes are obtained with $\left<I\right> \propto a^{2}$ thanks to the electric-magnetic duality). In addition to being phenomenologically interesting, this behaviour is motivated by a global attractor solution of the classical field equations, which exists for quite general classes of coupling functions $I(\phi)$ and inflaton potentials $V(\phi)$. This background solution possesses a homogenous background vector $\mathbf E_0$ in an anisotropically inflating universe whose magnitude scales as $|\mathbf E_0| \propto H$, namely $\left<I\right> \propto a^{-2}$ when ignoring slow-roll corrections \cite{attractor1,attractor2,attractor3,attractor4,Yamamoto:multivector1,attractor5,attractor6}.\footnote{The expansion anisotropy can be tuned small by the parameters of the Lagrangian in agreement with the $\Delta H / H \sim 10^{-8}$ constraint discussed above.}  Apparently, then, scale invariance is attractive. However, in the next section we shall review how energy is being transferred from the ultraviolet to super-horizon scales via gauge modes which become classical upon horizon crossing. We acknowledge the fact that it is presently unknown whether the $\left<I\right> \propto a^{-2}$ attractor is stable under quantum back-reaction, but, nevertheless, stick to the same assumption as most of the literature cited in this paper (for an exception, see \cite{Nurmi:2013gpa} which considers a free parameter in the gauge field power spectrum).

\subsection{Generation of vector fluctuations \label{chvec}}
After quantization in a gauge with vanishing temporal component $A_0$, the (three) vector potential can be written $\mathbf A =\mathbf A_0 + \delta\mathbf A$, where $\mathbf A_0$ denotes the solution of the classical equations while the quantum fluctuations (see \cite{mukhanov,mukhanovsbook} for details on the quantization procedure) 
\be
\delta \mathbf A(\tau,\mathbf x) = \sum_{\lambda=\pm} \frac{\int d^3k}{(2\pi)^{3/2}} e^{i\mathbf k \cdot \mathbf x} \boldsymbol{\epsilon}_\lambda(\mathbf k) \frac{\hat V_\lambda(\tau,\mathbf k)}{\left<I\right>},
\label{Aexpanded}
\ee
are expanded in circular polarization vectors satisfying $\mathbf k \cdot \boldsymbol{\epsilon}_\pm(\mathbf k)=0$, $\mathbf k \times \boldsymbol{\epsilon}_\pm(\mathbf k)=\mp ik\boldsymbol{\epsilon}_\pm(\mathbf k)$ and $\boldsymbol{\epsilon}_\lambda^*(\mathbf k) \cdot \boldsymbol{\epsilon}_\lambda'(\mathbf k) = \delta_{\lambda\lambda'}$.
The operator is defined 
\be
\hat V(\tau,\mathbf k) = a_\lambda(\mathbf k) V_\lambda(\tau,k) + a_\lambda^\dagger(-\mathbf k) V_\lambda^*(\tau,k),
\ee
where the mode function is
\be
V_\lambda(\tau,k) = \frac{1+ik\tau}{\sqrt 2 k^{3/2} \tau} e^{-ik\tau},
\ee
and the annihilation / creation operators satisfy $\left[a_\lambda(\mathbf k), a^\dagger_{\lambda' }(\mathbf p)\right] =  \delta_{\lambda\lambda'}\delta^3(\mathbf k - \mathbf p)$. The corresponding fluctuations of the field strength can be decomposed in electric- and magnetic-type components
\be
\delta \vE = -\frac{\left<I\right>}{a^2} \delta \mathbf A', \qquad \delta \vB = -\frac{\left<I\right>}{a^2} \nabla \times \delta \mathbf A.
\label{decomposition}
\ee

Following \cite{peloso12} we let $\vE_{IR}$ and $\vB_{IR}$ denote the parts of $\delta \vE$ and $\delta \vB$ that only consist of superhorizon modes so that the integration in (\ref{Aexpanded}) is restricted to wave vectors satisfying 
\be
\mathcal H_i < k <\mathcal H(\tau).
\label{integrationlimit}
\ee
The lower limit is the value of $\mathcal H = a H$ at the beginning of inflation which eliminates modes that has never been touched by the inflationary machinery (which is a reasonable assumption since quantum fluctuations are generated in the ultraviolet). The upper limit cuts off modes smaller than the horizon. Notice that $\vE_{IR}$ ($\vB_{IR}$) has not the same time dependence as $\delta\vE$ ($\delta\vB$) since $\mathcal H$ (and therefore the upper limit of integral) is dynamical.  We shall disregard slow-roll corrections and assume that $H$ is a constant during inflation so that it is related to the conformal time parameter by $a H=-1/\tau$.

Inserting (\ref{Aexpanded}) into (\ref{decomposition}) and using the identities of the polarization vectors leads to 
\begin{align}
\vE_{IR}(\tau,\mathbf x) &=  \int_{IR} \frac{d^3k}{(2\pi)^{3/2}} e^{i\mathbf k \cdot \mathbf x} \delta\boldsymbol{\mathcal{E}}(\tau,\mathbf k), \label{EIR}\\
\vB_{IR}(\tau,\mathbf x) &=  \int_{IR} \frac{d^3k}{(2\pi)^{3/2}} e^{i\mathbf k \cdot \mathbf x} \delta\boldsymbol{\mathcal{B}}(\tau,\mathbf k),\label{BIR}
\end{align}
where
\begin{align}
\delta\boldsymbol{\mathcal{E}}(\tau,\mathbf k) &\simeq \sum_{\lambda=\pm}  \boldsymbol{\epsilon}_\lambda(\mathbf k) \frac{3H^2}{\sqrt{2} k^{3/2}} (a_\lambda(\mathbf k)+a_\lambda^\dagger(-\mathbf k)), \label{fourierE}\\
\delta\boldsymbol{\mathcal{B}}(\tau,\mathbf k) &\simeq \sum_{\lambda=\pm}   \lambda  \boldsymbol{\epsilon}_\lambda(\mathbf k)  \frac{H^2 k \tau}{\sqrt{2}k^{3/2} } (a_\lambda(\mathbf k)+a_\lambda^\dagger(-\mathbf k)). \label{fourierB}
\end{align}
As explained above the integration domain in (\ref{EIR})-(\ref{BIR}) is over all $\mathbf k$ satisfying (\ref{integrationlimit}). In (\ref{fourierE})-(\ref{fourierB}) we have neglected quadratic and higher order terms in $k\tau$ which is a good approximation for superhorizon modes. We note that $\vB_{IR}$ is decreasing and suppressed by the small prefactor $k\tau$ relative to $\vE_{IR}$.  Contributions from $\vB_{IR}$ to the primordial power spectrum can therefore be neglected and we shall mainly focus on $\vE_{IR}$ below.  

The transition from a quantum field to a classical vector field coincides more or less with horizon crossing (and that is so because of the source term: it would not happen in vacuum since the massless vector field is conformally invariant \cite{Tinyakov:2013,Albrecht:1992kf}). First we note that $\delta\boldsymbol{\mathcal{E}}$ and $\delta\boldsymbol{\mathcal{B}}$ commute at first order, with each other and with themselves (in the sense that $\delta\boldsymbol{\mathcal{E}}(\tau_1,\mathbf k)\delta\boldsymbol{\mathcal{E}}(\tau_2,\mathbf k) = \delta\boldsymbol{\mathcal{E}}(\tau_2,\mathbf k)\delta\boldsymbol{\mathcal{E}}(\tau_1,\mathbf k)$ where $\tau_1 \neq \tau_2$); this is not the case for subhorizon scales where higher order terms in $k\tau$ cannot be neglected. This is a consequence of the fact that the raising and lowering operator have the same time dependency in the large wavelength regime. For the same reason the eigenvectors of $\delta \boldsymbol{\mathcal{E}}$ do not change with time.  For all these reasons we can view $\vE_{IR}$ as a classical field.  Moreover, as pointed out in \cite{peloso12}, since it only consists of superhorizon modes, $\vE_{IR}$ looks like a homogeneous vector field to a local observer.  A homogenous classical vector field is one which points in the same direction and with the same magnitude across a spatial slice. It therefore introduces a preferred direction in the universe. Since local observers do not have access to the fluctuations of $\vE_{IR}$, an extreme case of cosmic variance is introduced. Such an observer will only be able to measure a single value $\vE_{IR}(\tau)$. This realization is drawn from a Gaussian probability distributions with variance given by the vacuum expectation value: 
\begin{align}
&\left<0| \vE_{IR} \cdot \vE_{IR} |0\right> = \frac{9H^4}{2\pi^2} N,
\label{exp1} 
\end{align}
where $N$ is the number of e-folds since the start of inflation ($a=a_ie^{N}$). Hence, the dynamics of $\vE_{IR}$ can be viewed as a Gaussian random walk with a new piece drawn from a distribution with variance $9H^4/2\pi^2$ added for each e-fold. We emphasize that, since the modes originate from quantum fluctuations in a FLRW background (to good approximation), the direction of the new piece is random and, consequently, the total vector rotates/precess under the stochastic build-up (we characterize the dynamics statistically in section \ref{chstats}). 

\subsection{Effective interaction Hamiltonian \label{chpower}}
Here we shall write down the essentials needed to calculate the curvature power spectrum including the effective interaction Hamiltonian associated with loop terms. 

We follow a procedure similar to the one in \cite{peloso12} and start by splitting the power spectrum in two pieces
\be
\mathcal P_\zeta (\mathbf k) = \mathcal P_0(k) + \delta \mathcal P(\mathbf k), 
\ee 
where the first term is the dominant isotropic part and the latter represents the anisotropic correction produced by the vector.\footnote{We shall see that $\delta \mathcal P(\mathbf k)$ is not a pure quadrupole, but also gives a (small) contribution to the monopole. Therefore $\mathcal P_0$, is not exactly the same as the isotropically averaged power spectrum $\mathcal{P}_\mathcal{I}$ in equation (\ref{formidable}).} For the monopole we are only interested in the lowest order (de Sitter) expression
\be
\mathcal P_0 = \frac{H^2}{8\pi^2\epsilon M_p^2}.
\ee  
The quadrupole correction is calculated via the in-in formalism
\be
\begin{split}
\delta\left< \hat\zeta_{\vk}\; \hat\zeta_{\vp} (\tau) \right> &= -\int_{\tau_\text{min}}^{\tau}d\tau_1\int_{\tau_\text{min}}^{\tau_1}d\tau_2 \left< \left[ \left[ \hat\zeta^{(0)}_{\vk} \; \hat\zeta^{(0)}_{\vp}, H_\text{int}(\tau_1) \right], H_\text{int}(\tau_2) \right] \right> \label{correlationfunction} \\
&= 2\pi^2 \frac{\delta^3(\mathbf k + \mathbf p)}{k^3} \delta \mathcal P(\mathbf k),
\end{split}
\ee
where the unperturbed FLRW operator is given by
\be
\hat\zeta^{(0)}_{\vk} = \zeta^{(0)}_{\vk} a_\mathbf{k} + \zeta^{(0)*}_{\vk} a_{-\mathbf{k}}^\dagger, \quad \zeta^{(0)}_{\vk} \simeq \frac{H(1+ik\tau)}{2\sqrt{\epsilon} M_p k^{3/2}}e^{-ik\tau},
\ee
and the effective interaction Hamiltonian associated with $\vE_{IR}$ is \cite{peloso12}
\be
H_\text{int}(\tau) = -\frac{4}{H^4\tau^4} \int d^3k \mathbf \; \left( \vE_{IR}(\tau) \cdot \delta\boldsymbol{\mathcal{E}}(\tau,\mathbf k) \right) \hat\zeta^{(0)}_{-\mathbf k}.
\label{interactionH}
\ee

There will be an additional contribution coming from the vacuum expectation value $\vE_0$ which represents the attractor solution of the classical equations of motion. This gives another term to the interaction Hamiltonian similar to (\ref{interactionH}) with $\vE_{IR} \rightarrow \vE_{0}$.  We shall mainly neglect the contribution from $\vE_0$ in this paper since it is both homogenous and stationary (it does not rotate), so its consequences are already well understood. 

The identification of the effective interaction Hamiltonian (\ref{interactionH}) is a bit non-trivial; we refer to \cite{peloso12} for the details and just provide an explanation of the essential argument here. It represents a loop correction originating from the $\delta \vE \cdot \delta \vE$ term in the interaction Lagrangian.  When taking this term directly into account, it leads to an isotropic correction to the power spectrum because $\delta \vE$ is drawn from a distribution that respects rotational invariance by assumption. The corresponding theoretical expectation value could only be tested empirically by a hypothetical super observer with access to the entire inflated space. However, as local observers we are limited by causality in our ability to test our models empirically. Specifically, the superhorizon part of $\delta \vE$, namely $\vE_{IR}$, violates isotropy since it looks like a homogeneous classical vector locally.  As our local patch of universe corresponds to a particular realization of $\vE_{IR}$, we are interested in the effective power spectrum describing such patches (which obviously must be described by stochastic parameters).  Therefore, the effective contribution from the loop term $\propto \delta \vE \cdot \delta \vE$ in a given realization is actually of the form $\propto \vE_{IR} \cdot \delta \vE$ which leads to the interaction Hamiltonian (\ref{interactionH}).

We shall finish this section by providing an expression for the quadrupole correction $\delta\mP (\vk)$ which allows for the full time dependence of $\vE_{IR}$, thence including its rotation, to be accounted for. Carrying out the commutation relations in the correlation function (\ref{correlationfunction}) we can write
\be
\begin{split}
\delta\left< \hat\zeta_{\vk}\; \hat\zeta_{\vp} (\tau) \right> = \frac{4}{9\epsilon^2M_p^4 H^4}\int_{\tau_\text{min}}^{\tau}d\tau_1\int_{\tau_\text{min}}^{\tau_1}d\tau_2 \left(\frac{\tau^3-\tau_1^3}{\tau_1^4}\right) \left(\frac{\tau^3-\tau_2^3}{\tau_2^4}\right) \\
\left<  \left( \vE_{IR}(\tau_1) \cdot  \delta\boldsymbol{\mathcal{E}}(\tau_1,\mathbf k)\right) \left( \vE_{IR}(\tau_2) \cdot  \delta\boldsymbol{\mathcal{E}}(\tau_2,\mathbf p)\right) + (\mathbf k \leftrightarrow \mathbf p) \right>.
\end{split}
\label{longexpression}
\ee
Using 
\be
\sum_\lambda \epsilon_{\lambda,i}(\mathbf k) \epsilon_{\lambda,j}^*(\mathbf k) = \delta_{ij}-\frac{k_ik_j}{k^2}
\ee
and the definition (\ref{fourierE}) of $\delta\boldsymbol{\mathcal{E}}$ the expectation value on the second line of (\ref{longexpression}) can be calculated leading to the power spectrum
\be
\begin{split}
\delta\mathcal P(\mathbf k) = \frac{48}{\epsilon} \frac{\mathcal P_0}{V(\phi)} \int_{-1/k}^{\tau_\text{end}}d\tau_1\int_{-1/k}^{\tau_1}d\tau_2 \left(\frac{\tau^3-\tau_1^3}{\tau_1^4}\right) \left(\frac{\tau^3-\tau_2^3}{\tau_2^4}\right)\\
\left[ \vE_{IR}(\tau_1) \cdot \vE_{IR}(\tau_2) - \left( \hat{\vk} \cdot \vE_{IR}(\tau_1)  \right) \left( \hat{\vk} \cdot \vE_{IR}(\tau_2)  \right)  \right].
\label{powerComplicated}
\end{split}
\ee
The contribution from times where a mode is sub horizon is negligible so we have set $\tau_\text{min}=-1/k$ as lower limit in the integrals to ensure that only super horizon fluctuations are taken into account.\footnote{This lower cutoff also ensures that the initial conditions for the original path integral representation of this two-point function are properly accounted for in this form, see \cite{Adshead:2009cb}.}

\section{The IR Landscape (Bubbland) \label{chbubble}}

As explained above, the effective interaction Hamiltonian under consideration depends explicitly on $\vE_{IR}$, namely the sum of gauge modes which at a given instant are super horizon.  Although $\vE_{IR}$ is a quasi-homogeneous vector, pointing in a fixed direction and with a constant magnitude over the accessible spatial patch, it varies significantly over length scales much larger than the horizon. This leads to a landscape of associated signatures where local observables depend on the position of the observer within the entire inflated space. Hence the causally connected patch of the universe, Bubbland, which very likely is exponentially larger than our observable universe (as is the case if inflation lasted even only a few e-folds longer than what is required to solve the horizon problem), consists of a multitude of bubbles each with its own signatures dictated by the local realization of the infrared vector.  

The contribution from $\vE_{IR}$ to the primordial power spectrum $P_\zeta (\mathbf k)$ was first taken into account in \cite{peloso12}. See also \cite{dimopoulos} for a study of the inflationary buildup of infrared modes in more general classes of models with light vector spectator fields. However, as we shall study quantitatively in section \ref{chstatistics}, $\vE_{IR}$ possesses a rotation which is significant over the time interval the CMB modes cross the horizon. This phenomenon has so far been neglected in the literature and one of the main goals of this section is to discuss its implications for $P_\zeta (\mathbf k)$ and the interpretation thereof. We shall also see that the intrinsic randomness of the infrared vector transfers directly over to local observables: we then use this to set constraints on the model in a statistically sound manner.

\subsection{The effective IR cutoff \label{chircut}}

Let us practically compute the time integrals appearing in (\ref{powerComplicated}).  Apparently, because of the (randomly) rotating vector $\vE_{IR}(\tau)$, $\delta\mathcal P$ has the form of a general quadrupole without any axial symmetry and should be parametrized as (\ref{scaleinvariant}).  This would be faulty, however, since $\vE_{IR}$ is not a truly homogenous vector.  Indeed, the only reason $\hat{\vE}_{IR}(\tau)$ has to rotate \emph{after} a given curvature perturbation with wavelength $1/k$ has crossed the horizon, is that new vector fluctuations $\delta\boldsymbol{\mathcal{E}}$ with wavelengths \emph{smaller} than $1/k$ become infrared and are added to $\vE_{IR}$ defined in (\ref{EIR}). Despite being super horizon, these new vector modes cannot be viewed as background for the considered curvature perturbation mode and should be subtracted off from $\vE_{IR}$.  A first approximation (justified in detail below) would be to subtract from $\vE_{IR}$ any mode with wavelengths smaller than our presently observable patch of the universe $1/\mathcal{H}_0$: in this way we make sure that we are dealing with homogeneous vectors from our point of view, and we obtain an effective power spectrum describing our (present) Hubble patch.  We thence opt for substituting $\vE_{IR}(\tau_1),\vE_{IR}(\tau_2)\rightarrow \vE_{IR}(\tau_0)$ in (\ref{powerComplicated}) where $\tau_0= - 1/\mathcal{H}_0$ is a constant under the integration.  In practice:
\begin{align}
\delta P(\mathbf k) &=  \frac{48}{\epsilon}\mathcal P_0 \frac{|\vE_{IR}(\tau_0)|^2}{V(\phi)} \left[1-\left(\hat{\vk}\cdot \hat{\vE}_{IR}(\tau_0) \right)^2\right]  \int_{-1/k}^{\tau_\text{end}}d\tau_1\int_{-1/k}^{\tau_1}d\tau_2 \left(\frac{\tau^3-\tau_1^3}{\tau_1^4}\right) \left(\frac{\tau^3-\tau_2^3}{\tau_2^4}\right) \nonumber \\ 
&\simeq\frac{24}{\epsilon}\mathcal P_0 \frac{|\vE_{IR}(\tau_0)|^2}{V(\phi)} N_k^2  \left[1-\left(\hat{\vk}\cdot \hat{\vE}_{IR}(\tau_0) \right)^2\right], \label{ps}
\end{align}
where $\hat{\vE}_{IR}=\vE_{IR} / |\vE_{IR}|$ is the normalized classical vector, and $N_k$ is the remaining number of e-folds of inflation when the comoving mode $k$ crossed the horizon.

This approximation neglects the effects of the gauge modes which lie in the range  1 Mpc $\lesssim 1/k \lesssim$ 1 Gpc; these CMB-accessible modes do have a small impact on the quadrupole correction which can be roughly quantified, to validate our approximation above.  During inflation all modes with $k > -1/\tau_{in}$ experience rotation.  That is because from $\tau_{in}$ to some given $\tau = -1/k$ more and more modes have been added to the $\vE_{IR}$ background.  Of course, as we have noticed before, anything which is beyond this $\tau$ makes no difference for $k$, but will make a difference for higher $k$.

\begin{figure}[tbph]
\begin{center}
\includegraphics[width=1.0\textwidth]{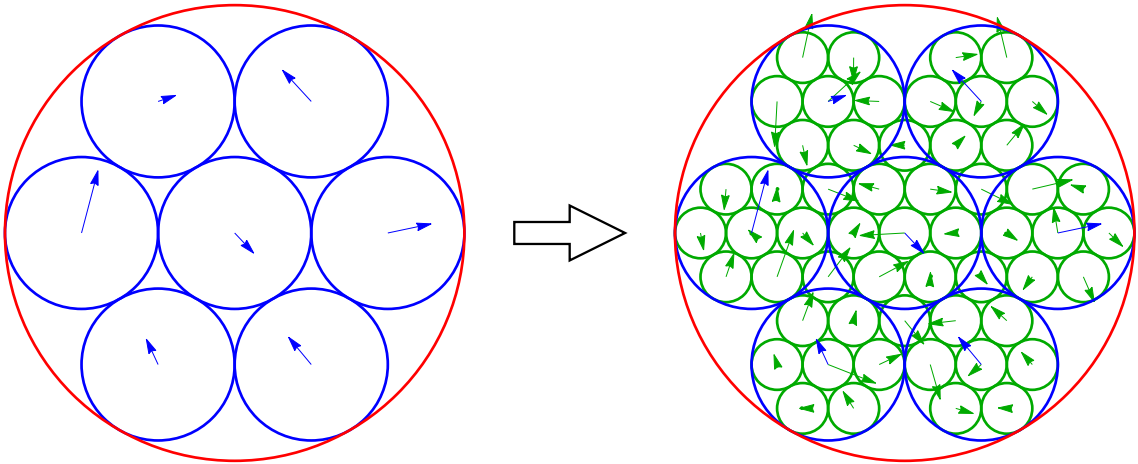}
\end{center}
\caption{Two dimensional schematic illustration of the evolution of a comoving patch with size equal to the present horizon.  The red circle represents the present observable patch $1/\mathcal{H}_0$ which left the horizon $N_\text{ex}$ e-folds after the start of inflation. The blue and green circles represents the horizon roughly $N_\text{ex}+1$ and $N_\text{ex}+2$ efolds after start of inflation, respectively. The blue and green arrows represents $\vE_{1}$ and $\vE_{2}$, namely the contribution to $\vE_{IR}$ added in the time intervals $(N_\text{ex},N_\text{ex}+1)$ and $(N_\text{ex}+1,N_\text{ex}+2)$, respectively. In this illustration $\vE_{1}$ and $\vE_{2}$ in different bubbles are drawn from (independent) normal distributions whereas in the real world neighbouring bubbles are correlated. However, the number of bubbles grows exponentially fast so quickly the typical distance between two (random) bubbles is so large that the correlation is negligible.}
\label{bblz}
\end{figure}

Let us look at the 1 Mpc mode.  During inflation, up until this mode leaves, the background $\vE_{IR}$ for this mode has been changing ever since the beginning; the background vector stops precessing (from the point of view of the 1 Mpc mode observer) as soon as this mode becomes superhorizon.  Since our present observable universe is much larger than this mode (which has fallen back into the horizon in the past), what we observe is a collection of several bubbles for each of which the rotation experienced by the $k = 1/$Mpc mode is random.  Then in general the quadrupole parameters which describe one bubble are not the same at those describing any other bubble. But the orientations are random, and there is a finite number of bubbles.  Thence, the observed actual rotation will be averaged out amongst the multitude of bubbles, see figure \ref{bblz}.

To quantify this suppression we proceed as follows.  For each e-fold a new piece $\vE_{n}$ with variance $\sim H^4$ is added to $\vE_{IR}$.  $N$ e-folds after $\mathcal{H}_0$ crossed the horizon there are $n_B=e^{3N}$ bubbles.  The distribution of the average of $n_B$ independent normal distributions with variance $H^4$ is another normal distribution with variance $H^4/n_B$. Thus the variance of the new contribution is suppressed by a factor $e^{-3N}$.  This demonstrates that the corrections are very small and that the quadrupole will be axisymmetric to a very good approximation, namely, $\chi\simeq0$ and the power spectrum can be parametrized by the usual $g$ parameter alone.  Of course, it is not strictly true that each bubble is independent, especially neighboring bubbles are highly correlated. However, when N grows, the distance between two random bubbles is typically much larger than the wavelength of the new modes which have recently crossed the horizon (again, see figure \ref{bblz}), so that the correlation is extremely low and our estimate becomes more precise. To summarize: in principle the precession affects the primordial power spectrum also in the single vector case. That effect however is tiny (and totally negligible if the total duration of inflation is significantly longer than the $60$ or so required to solve to horizon problem), which justifies our steps leading to (\ref{ps}). Of course this clear boundary with a cut-off at $k=\mathcal{H}_0$ is not so sharp in the real world, but makes the analysis very tractable, intuitive, and mostly analytic.

The total power spectrum can thus be written in the standard form (\ref{planarsymmetric})
\be
\mathcal P_\zeta (\mathbf k) = \mathcal P_0 \left(1+g(k) \cos^2 \vartheta_{\hat{\vk}, \hat{\mathbf n}}\right),
\label{standard}
\ee  
where 
\begin{align}
\hat{\mathbf n} &= \hat{\vE}_{IR}(\tau_0), \label{nstar} \\
g(k) &= - \frac{24}{\epsilon} \frac{|\vE_{IR}(\tau_0)|^2}{V} N_k^2,
\label{gstar}
\end{align}
and $\vartheta_{\hat{\vk}, \hat{\mathbf n}}$ denotes the angle between the wave vector $\mathbf k$ and the preferred direction $\hat{\mathbf n}$. Here we have employed the fact that $g$ is constrained observationally to be much less than unity to simplify its expression. Mathematically this is identical to the result in \cite{peloso12}, although this pioneering paper did not discuss the rotation of $\vE_{IR}$ and omitted the complications we focus on here. What we would like to point out is that (\ref{standard}) represents an effective power spectrum which is observer-dependent. By an `observer' we here mean a clever cosmologist able to reconstruct the primordial power spectrum from cosmic data. Specifically, the quantities $\hat{\mathbf n}$ and $g$ are dictated by the status of $\vE_{IR}$ at the time $\tau_0=-1/\mathcal{H}_0$ when the mode corresponding to the horizon of the observer crossed the horizon. This implies that neither the direction nor the amplitude of the quadrupole are fixed. Two observers living at two different times, but at the same (comoving) position, will reconstruct different amplitudes and directions.

As an example, consider an observer at redshift $z=6$. His comoving horizon is a factor $e$ smaller than the present horizon, $(1/\mathcal H)_{z=6} \sim 1/(e \mathcal{H}_0)$. The classical vector $\vE_{IR}$ dictating the imprints in this patch is therefore not the same as the vector creating imprints in our (larger) patch. Over one e-fold of inflation $\vE_{IR}$ has both rotated and changed magnitude, and the quadrupole's amplitude and shape imprinted in cosmic data are going to be (statistically) different. Thus the two observers cannot expect to agree on the preferred direction they observe on the sky although they inhabitate the same position and are not that much separated in time. 

A comment now on the role of the background vector attractor solution $\vE_0$, which we have neglected here.  As mentioned above it gives a contribution to the interaction Hamiltonian similarly to $\vE_{IR}$.  The total power spectrum therefore takes the same form as (\ref{standard}) with the substitution $\vE_{IR}\rightarrow \vE_\text{classical}= \vE_{0} + \vE_{IR}$ \cite{peloso12}. Although the contribution from $\vE_{IR}$ is identical to that of $\vE_0$ mathematically, its interpretation is quite different.  Unlike the sum of the infrared fluctuations, namely $\vE_{IR}$, the background attractor solution $\vE_0$ is truly homogeneous: in a sense it corresponds to an infinite wavelength. There is therefore no inhomogeneous part to cut off and, consequently, the contribution from $\vE_0$ is not observer dependent (and it does not precess); an observer at $z=100$ or in the future would reconstruct its impact on the primordial power spectrum in the same way as an observer today.

\subsection{Phenomenology \label{chplanck}}
Here we shall constrain the total duration of inflation (assuming existence of the gauge kinetic coupling) adopting the recent analysis of \cite{komatsu} which put the model independent constraint $0.002\pm0.016$ ($68\%$ CL) using Planck data and the analysis of \cite{pelosoconstraint} yielding $g_0<0.02 \;(95\%$ CL) for models with the $I^2F^2$ interaction. We shall restrict our attention to the case of a single vector since bounds for multi-vector models were already reported in \cite{Thorsrud:2013mma}.

Inserting $N_{\mathcal{H}_0} = 60$ together with the slow-roll value $\epsilon=0.01$ in (\ref{gstar}), we find that $g_0\equiv g(\mathcal{H}_0)=-0.1$ corresponds to $|\vE_{IR}(\tau_0)|^2 / V \sim 10^{-8}$. Hence, a $10\%$ anisotropy is produced by a vector field which is suppressed $8$ orders of magnitude compared to the inflaton energy density. This well-known result corroborates our choice of isotropic FLRW metric.  Next, we use it to set $3H^2 M_p^2/V=1$ so that (\ref{gstar}) can be written 
\be
g_0 \simeq -\kappa^2 N_\text{ex} \tilde{\vE}^2,  \qquad \kappa^2 = 96 \mP_0 N_{\mathcal{H}_0}^2, 
\label{phenomenology}
\ee
where $\tilde{\vE} = \vE_{IR}(\tau_0)/ \sqrt{3H^4N_\text{ex}/ 2\pi^2}$ is a rescaled vector having the expectation value $\langle \tilde{\vE}^2 \rangle = 3$ according to (\ref{exp1}). Here $N_\text{ex}$ is the ``extra'' e-folds of inflation in addition to the $60$ or so needed to solve the horizon problem. With this redefinition the stochastic component of $g_0$ is encoded in $\tilde{\vE}^2$ whereas the dependence on the total duration of inflation is given explicitly by the factor $N_\text{ex}$ in (\ref{phenomenology}).  Each component of  $\tilde{\vE}$ is drawn from a Gaussian distribution with unit variance and $g_0$ is proportional to the chi-square distributed variable $\tilde{\vE}^2$. It follows that the probability distribution function (PDF) of the amplitude of the quadrupole is
\be
\mathbf{P}(g_0) = \frac{2\pi}{(2\pi\kappa^2 N_\text{ex})^{3/2}} \sqrt{-g_0} \exp{\left[ \frac{g_0}{2\kappa^2N_\text{ex}} \right]}.
\label{phPDF}
\ee 

\begin{figure}[tbph]
\begin{center}
\includegraphics[width=0.7\textwidth]{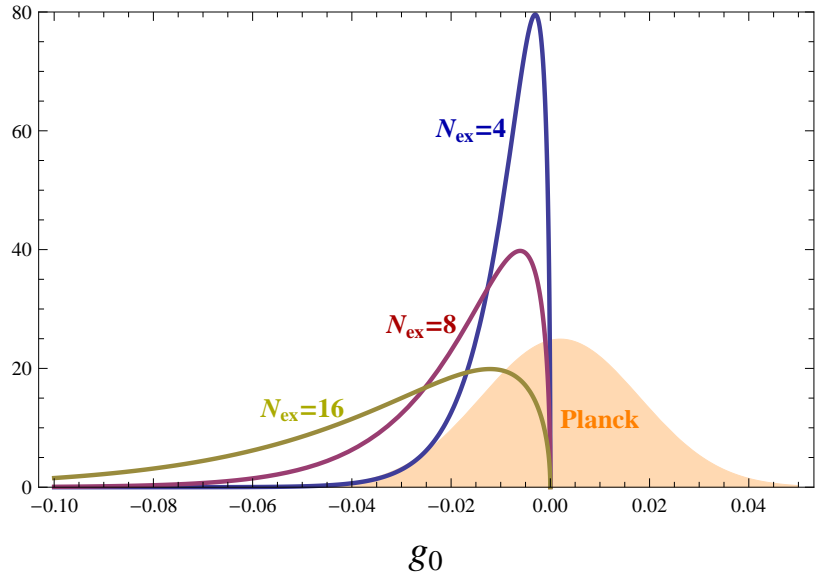}
\end{center}
\caption{Probability distribution functions of $g_0$ for three examples with $N_\text{ex}=\{ 4,8,16 \}$ in models with a single gauge vector. The orange region corresponds to the distribution inferred from the Planck constraints reported in \cite{komatsu}.}
\label{ph1}
\end{figure}

The probability for $g_0>0$ is zero and the function is normalized so that integration over all negative reals equals unity.  For the unperturbed spectrum we can use the Planck best-fit value $\mathcal P_0 \simeq \mathcal P_\zeta \simeq 2.2 \cdot 10^{-9}$ \cite{planck}. The largest observable mode left the horizon around $60$ e-folds before the end of inflation in canonical models so, for concreteness, we set $N_{\mathcal H_0}=60$. However, the PDF also depends on the free parameter $N_\text{ex}$; in figure \ref{ph1} we have plotted the PDFs for three examples with $N_\text{ex}=\{4,8,16\}$ together with the Planck distribution/constraint \cite{komatsu}.\footnote{The limits reported in \cite{komatsu} are $0.002\pm0.016$ ($68\%$ CL), $0.002^{+0.031}_{-0.032}$ ($95\%$ CL) and $0.002^{+0.037}_{-0.048}$ ($99.7\%$ CL). This indicates that the distribution is (very) close to a Gaussian with mean $0.002$ and standard deviation $0.016$ which is the distribution (orange region) drawn in figure \ref{ph1}. Note that we have not shown the distribution corresponding to the model-specific limit reported in \cite{pelosoconstraint}. The reason for this is that it relies on a fitting-function (which is non-Gaussian, see figure 19 in \cite{PlanckNonGauss})) whose explicit form is unknown to us.} 

\begin{figure}[tbph]
\begin{center}
\includegraphics[width=0.7\textwidth]{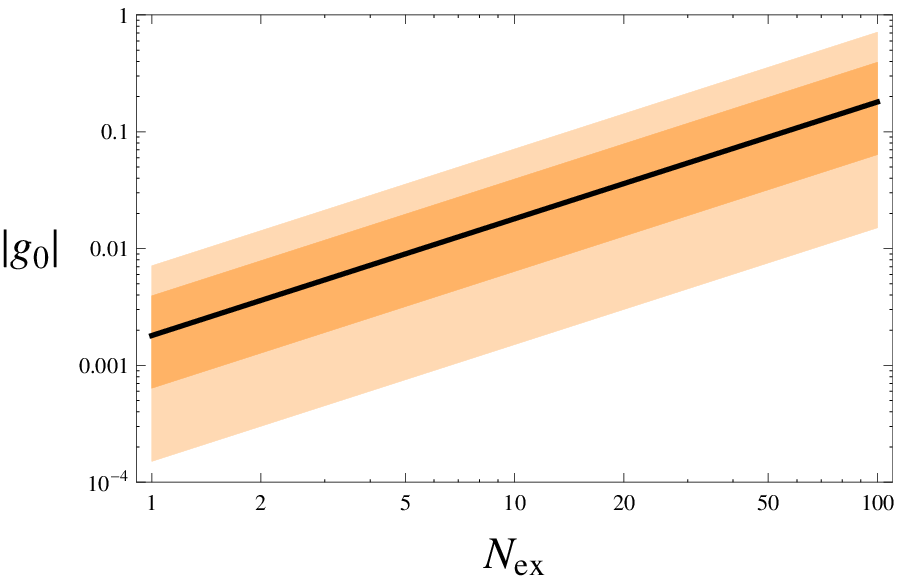}
\includegraphics[width=0.7\textwidth]{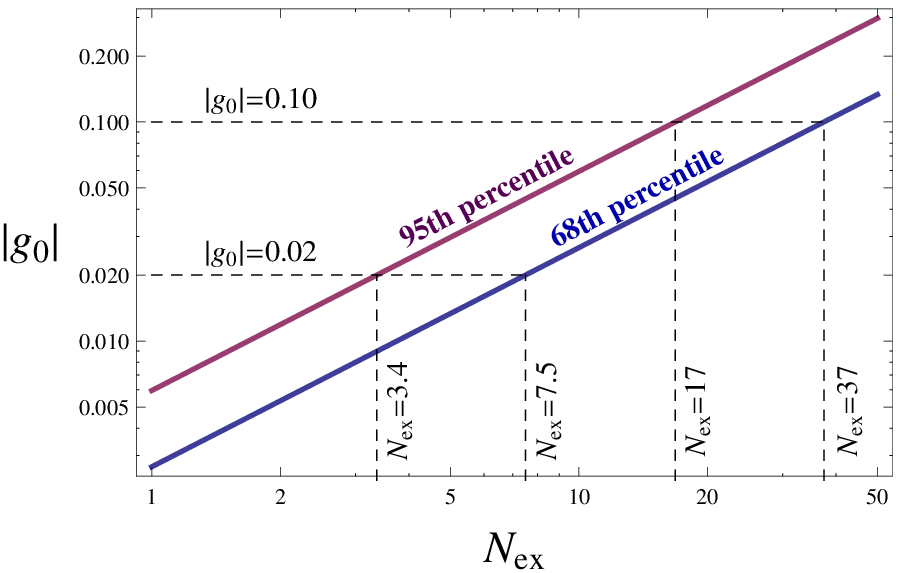}
\end{center}
\caption{Upper panel: median value (black line) and $68\%$ ($95\%$) confidence interval in orange (light orange) of $|g_0|=-g_0$. Lower pandel:  68th and 95th percentiles of $|g_0|$ as a function of $N_\text{ex}$. Intersections with $|g_0|=0.02$ and $|g_0|=0.10$ are indicated explicitly.}
\label{ph2}
\end{figure}

We are then ready to constrain the free parameter phenomenologically.  Per definition, integration of $\mathbf{P}(g_0)$ from the $i$-th percentile to $0$ equals $i/100$. In the upper panel of figure \ref{ph2} we show the median value of $|g_0|=-g_0$ together with the $68\%$ and $95\%$ confidence intervals as functions of $N_\text{ex}$ (build from the $2.5$th, $16$th, $84$th, $97.5$th percentiles).  Note that $g_0$ varies as much as one (two) order(s) of magnitude over the $68\%$ ($95\%$) regions. In the lower panel of figure \ref{ph2} we have plotted the $68$th and $95$th percentiles as functions of $N_\text{ex}$. We have explicitly marked the intersections with $|g_0|=0.02$ and $|g_0|=0.10$. For instance, if $N_\text{ex}=7.5$ there is $68\%$ chance for observing $|g_0|<0.02$, or conversely, $32\%$ chance for $|g_0|>0.02$. At $95\%$ confidence the similar statistics is obtained for $N_\text{ex}=3.4$. In a similar way constraints on $N_\text{ex}$ for the looser limit $g_0<0.10$ can be read off from the figure.  

We conclude that if $N_\text{ex}$ is around 8 (4) or larger, the IR fluctuations alone creates a quadrupole with expectation values in conflict with constraints reported in \cite{komatsu} (\cite{pelosoconstraint}).  Our reported bounds are conservative in the sense that we disregarded the contribution from the background vector $\mathbf E_{0}$, which would increase the expectations for the total quadrupole (after integration over all possible alignments between $\vE_0$ and $\vE_{IR}$).\footnote{$|\mathbf E_{0}|$ is essentially is a free parameter determined by the parameters of the interaction Lagrangian and inflaton potential.}  However, it should be emphasized that these limits are obtained under the assumption $\left< I \right>\propto a^{-2}$ whose validity depends on the (unknown) dynamical effect of quantum back-reaction as discussed in section \ref{chmodel}. Taken at face value our derived tight limits on $N_\text{ex}$ suggest that the largest observable modes might not have reached the vacuum state upon horizon crossing and may carry information from the pre-inflationary phase as discussed in \cite{Chen:2013tna} for a minimally coupled Maxwell vector and in \cite{Emami:2014tpa} for the $I^2(\phi)F^2$ coupling in particular.

\subsection{Collection of vectors, and non-Gaussianity \label{3pf}}
Above we justified that, as an approximation, the power spectrum can be written in the axisymmetric form (\ref{planarsymmetric}).  To demonstrate the full capability of the parametrization developed in section \ref{chparameterizations}, we shall here briefly consider the multivector case where the two parameters ($g,\chi$) go hand in hand; for a complete presentation of the statistical predictions and implications see our previous work \cite{Thorsrud:2013mma}.  We also report the expressions for the three-point functions for the case of multiple gauge fields with uniform coupling at the end of this section.

Let $A^{(i)}_\mu$ denote $n$ copies of the Abelian gauge field. The interaction Lagrangian (\ref{lag}) generalizes to
\be
\mathcal{L}_\text{int} = \sum_{i=1}^n  -\frac{1}{4} I^2(\phi) F^{(i)}_{\mu\nu} F^{(i)\mu\nu},
\ee
where $F^{(i)}_{\mu\nu} = \nabla_\mu A^{(i)}_\nu - \nabla_\nu A^{(i)}_\mu$.  The interaction Hamiltonian (\ref{interactionH}) generalizes in an obvious way: since operators of different gauge fields commute, the quadrupole correction (\ref{ps}) becomes
\be
\begin{split}
\frac{\delta \mathcal P(\mathbf k)}{\mathcal P_0} &= \sum_{i=1}^n  \frac{24}{\epsilon}  \frac{(\vE_{IR}^{(i)})^2}{V(\phi)} N_k^2  \left[1-\left(\hat{\vk} \cdot \hat\vE_{IR}^{(i)} \right)^2\right] \\
&\equiv a_{00} Y_{00} + \sum_{m=-2}^2 a_{2m} Y_{2m}(\hat{\vk}),
\label{ps3}
\end{split}
\ee
which, barring special alignments between the different vectors, can in general not be recast in the axisymmetric single-vector form. The collection of vectors contributes with an isotropic piece, encoded in the spherical harmonic component $a_{00}$, as well as the anisotropic part (encoded in $a_{2m}$). The total power spectrum can then be written on the form
\be
\mathcal P(\mathbf k) = \mP_0(1+a_{00}Y_{00}) \left[ 1+ \frac{\Sigma_m a_{2m}Y_{2m}}{1+a_{00}Y_{00}}  \right],
\label{heip}
\ee
where $\mP_0(1+a_{00}Y_{00})$ is the isotropically averaged power spectrum (compare to (\ref{Qnotrotated})).

We are interested in the case that the dominant contribution to the power spectrum comes from the scalar field, $a_{00}\ll1$.  The smallness of the anisotropic modulation is, as usual, controlled by the amplitude $g$. In the single vector case $|a_{00}|\sim |a_{2m}|$ so that the smallness of $|g|$ guarantees the smallness of $|a_{00}|$. In the multivector case, however, this is not the case because $a_{00}$ scales proportionally to $n$ whereas $a_{2m}$ scales as $\sqrt{n}$ (verified below). The condition $a_{00}\ll1$ leads to the additional constraint \cite{Thorsrud:2013mma} 
\be
nN_{ex} \ll 10^4,
\label{condition1}
\ee 
in which case the total spectrum (\ref{heip}) takes the form (\ref{Qnotrotated}) with $\mP_0\simeq\mathcal P_\mathcal{I}$ and $a_{2m} \simeq b_{2m}$. Let us check how this condition compares to our assumption that the vectors are spectator fields, namely that the total energy budget is dominated by the (slowly rolling) inflaton: $\sum_{i=1}^n |\vE^{(i)}_{IR}|^2\ll V(\phi)$. Using (\ref{exp1}) we find $|\vE^{(i)}_{IR}|^2 \sim \langle |\vE^{(i)}_{IR}|^2 \rangle = 9H^4N_\text{ex}/2\pi^2$ so that the constraint reads
\be
nN_\text{ex} \ll \frac{1}{12\epsilon \mP_0} \simeq 10^9.
\ee  
Hence the spectator condition is much weaker than (\ref{condition1}). Before proceeding, we would like to point out that there exist a large regime $10^4 \ll n N_\text{ex} \ll 10^9$, where the total energy is dominated by the inflaton, whereas the curvature perturbations are dominated by the vectors (resembling curvaton models for instance). In that case (\ref{heip}) becomes
\be 
\mP=\mP_0 a_{00} Y_{00} \left[ 1 + \frac{\Sigma_m a_{2m} Y_{2m} }{a_{00}Y_{00}}  \right].
\ee
Since the power spectrum isotropizes as $g\propto a_{2m}/a_{00}\propto \sqrt{n}/n$ in this regime, it could be an interesting direction for further studies.   

\begin{figure}[tbph]
\begin{center}
\includegraphics[width=1.0\textwidth]{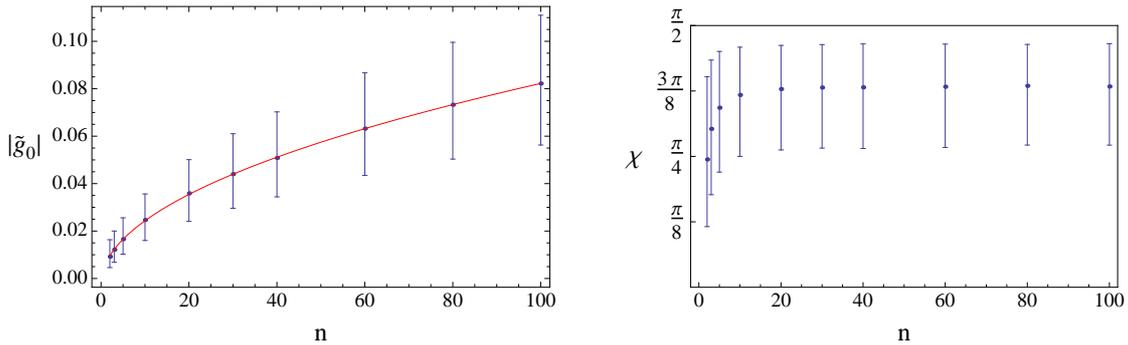}
\end{center}
\caption{Mean value of $|\tilde g_0| \equiv |g_0|/(N_\text{ex}/3)$ and $68\%$ error bars against number of gauge fields $n$. The red curve is the best fit function $10^{-3}(-2.4 + 8.5\sqrt{n})$.}
\label{figGandChi}
\end{figure}

From our formulas a Monte Carlo routine can be implemented to infer statistical predictions for the power spectrum.  By making realizations of the $n$ uncorrelated vectors $\vE^{(i)}_{IR}$, the spherical harmonic components, which under the condition (\ref{condition1}) takes the form
\be
b_{2m} = \int d\Omega \frac{\delta \mP}{\mP_0} Y_{2m},
\ee
can be rotated to the form $\tilde b_{2m}=(0,0,\tilde b_{20},\tilde b_{21},0)$ using the matrices in Appendix \ref{chproof}. From here the amplitude $g$ and shape $\chi$ characterizing the power spectrum (\ref{scaleinvariant}) are obtained via (\ref{gdef})-(\ref{chi2def}). In figure \ref{figGandChi} we show the mean value and $68\%$ error-bars as a function of number of gauge fields $n$ for $10^4$ realizations of $\vE^{(i)}_{IR}$. This verifies that the amplitude $g$ (and therefore $a_{2m}$) scales proportionally to $\sqrt n$ as claimed above.  


It is straightforward to go one step further and compute also the three-point function.  The single-vector case has been already discussed in \cite{peloso12}, so we refer to that paper for the technical steps of the computation, with the \emph{proviso} that once again in general the IR component $\vE_{IR}$ depends explicitly on time, so before taking the (three) time integrals one needs to specify that $\vE_{IR}$ is evaluated at $\tau\rightarrow\tau_0=-1/\mH_0$.

The multifield case can be again computed by a straightforward generalization of this result, leading to
\begin{align}\label{3pfmulti}
\delta\left< \hat\zeta_\vk\;\hat\zeta_\vp\;\hat\zeta_\vq (\tau) \right> =& \frac{27H^2}{\sqrt2\pi^{3/2}\epsilon^3M_p^6} \sum_{i=1}^n \left[ \left| \vE_{IR}^{(i)} \right|^2 \; \mathfrak{Kos}(\hat\vp,\hat\vq,\hat\vE_{IR}^{(i)}) \right] \\
& \left[ \prod_{j=1}^3 \int_{\tau_\text{min}}^{\tau}d\tau_j \left(\frac{\tau^3-\tau_j^3}{\tau_j^4}\right) \right] \frac{\delta^3(\vk+\vp+\vq)}{p^3q^3} \;+\; 2 \; \text{perm},
\end{align}
where the two permutations are obtained by rotating $\vk\rightarrow\vp\rightarrow\vq$, and the angular function $\mathfrak{Kos}$ is defined, with all $\vE_{IR}$ taken at $\tau_0$, as
\be
\mathfrak{Kos}(\hat\vp,\hat\vq,\hat\vE_{IR}^{(i)}) \equiv 1 - \left(\hat\vp \cdot \hat\vE_{IR}^{(i)} \right)^2 - \left(\hat\vq \cdot \hat\vE_{IR}^{(i)} \right)^2 + \hat\vp \cdot \hat\vq \; \hat\vp \cdot \hat\vE_{IR}^{(i)} \; \hat\vq \cdot \hat\vE_{IR}^{(i)}.
\ee
Once again, in general this expression can not be reduced to the single-field case (which can be obtained for $n=1$).

\section{Priors for cosmic data sets \label{chstatistics}}

In order to make definite and quantitative predictions capable of linking the fundamental theory to observations via different types of cosmological data sets we need to understand the statistical properties of the vector condensate. This is so because the correlators under consideration are dictated by the status of the IR vector(s) at the ``cut-off'' time $\tau_\text{cut}= -1/k_\text{cut}$, where $1/k_\text{cut}$ is the radius of the patch being probed. If the relevant cut-off is the causal horizon of the observer we have (roughly) $\tau_{\text{cut}} = -1/\mH_0$ as discussed in section \ref{chircut}. However, in practice the cut-off might not be set by causality alone. Whereas the CMB probes our entire Hubble patch, large scale structure (LSS) surveys are limited to redshifts below or around unity. Thence parameter reconstructions based on galaxy surveys are not expected to agree with those based on CMB. 
This deviation is unrelated to measurement uncertainties or systematic errors and would be present even if primordial correlators could be perfectly reconstructed from cosmic data. 
Statistically one expects a stronger anisotropy in LSS data because the comoving patch is smaller and therefore the cut-off more reductive (thereby typically producing a stronger IR vector). Another way to see it is that the CMB bubble contains several LSS bubbles.  Since the preferred direction in the smaller bubbles are different, the signal in the larger bubble is typically weaker (due to averaging effect). This is an interesting observational consequence of embedding fundamental or effective horizons in a bath of (classicalized) quantum fluctuations. We will study the effect at the quantitative level below and start by characterizing the dynamics of the IR vector.  

\subsection{Dynamics of the vector condensate: Gaussian random walk \label{chstats}}

In this section we shall characterize the dynamics of the IR vector statistically. Each component of $\vE_{IR}$ is drawn from a Gaussian probability function
\be
\mathbf{P}(E^{IR}_i) = \frac{1}{\sqrt{2\pi}\sigma} \exp{\left[-\frac{(E^{IR}_i)^2}{2\sigma^2}\right]},
\label{probability1}
\ee
with a variance that is one third of the expectation value in (\ref{exp1}):
\be
\sigma^2 = \frac{3H^4}{2\pi^2} N, \quad \left< E^{IR}_i E^{IR}_j \right> = \delta_{ij} \sigma^2.
\ee
It follows that the norm $E=|\vE_{IR}|$ is given by the chi distribution 
\be
\mathbf{P}(E) = \frac{4\pi E^{2}}{(\sqrt{2\pi}\sigma)^3} \exp\left[ -\frac{E^{2}}{2\sigma^2} \right],
\label{chisquared}
\ee
which is normalized according to $\int_0^\infty \mathbf{P}(E)dE=1$.

\begin{figure}[tbph]
\begin{center}
\includegraphics[width=1.0\textwidth]{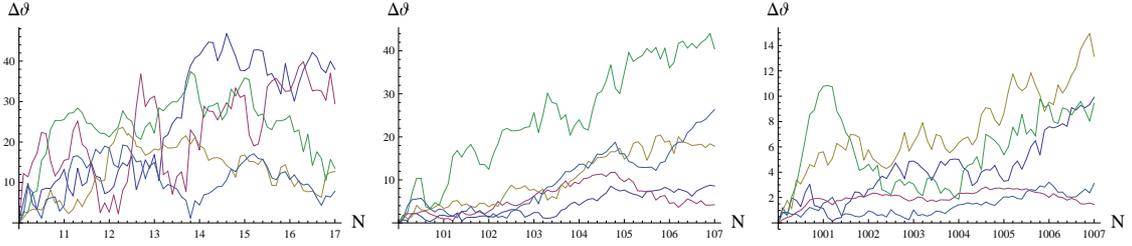}
\end{center}
\caption{The time evolution of the angular coordinate $\vartheta=\Delta \vartheta$ (measured in degrees) over 7 e-folds for random realizations of the Gaussian statistics associated with the vector. In the left-most figure inflation has lasted 10 e-folds before the start of the plot, while in the middle and right-most figure 100 and 1000 e-folds, respectively. For each case we show $5$ Montecarlo realizations. We note how the precession is suppressed in the panels to the right where the vector has had a longer time to build up (before the start of the plot). We have used 10 steps per e-fold and used a linear interpolation between the points.}
\label{fig:realizations}
\end{figure}

The variance is time dependent through its linear dependence on $N$, which represents the number of e-folds since the start of inflation. $\vE_{IR}$ is therefore a dynamical vector which rotates / precess and with a mean norm scaling as $\left<|\vE_{IR}|\right>\propto N$.  We can view the dynamics as a Gaussian random walk where a new contribution $ E^{\text{new}}_i$ (drawn from a Gaussian distribution) with variance $3H^4/2\pi^2$ is added to $E_i^{IR}$ for each e-fold. Figure \ref{fig:realizations} shows Monte Carlo simulations for the time evolution of the angular coordinate $\vartheta$ over 7 e-folds for 15 (random) realizations.     

We would like to know the typical precession and change in magnitude between two arbitrary times $N^a$ and $N^b$ with $N^b>N^a$ (we use the e-fold number $N$ as our clock in this section). We let $\vE^a\equiv\vE_{IR}(N^a)$ denote the realization at a time $N^a$ and $\vE^b$ at a later time $N^b$.\footnote{We shall always write $a$ and $b$ as superscripts since subscripts are reserved for the components of the vector. For instance $\vE^a = (E^a_x,E^a_y,E^a_z)$.}  Furthermore we let $\sigma_a^2$ and $\sigma_b^2$ denote the variances at these times, for instance $\sigma_a^2 = (3H^4/2\pi^2) N^a$ and similarly for $\sigma_b^2$.  The new field added during this time interval $\vE_\text{new}=\vE^b-\vE^a$ is drawn from the distribution (\ref{probability1}) with $\sigma^2$ replaced by 
\be
\sigma^2_\text{new}=\sigma_b^2-\sigma_a^2. 
\ee
It follows that the conditional probability density for $\vE^b$ given $\vE^a$ is
\be
\mathbf{P}(E^b_i|E^a_i) = \frac{1}{\sqrt{2\pi}\sigma_\text{new}} \exp{\left(-\frac{(E^b_i-E^a_i)^2}{2\sigma_\text{new}^2}\right)}
\ee
for each component, and thus 
\be
\mathbf{P}(\vE^b | \vE^a) = \frac{1}{(\sqrt{2\pi}\sigma_\text{new})^3} \exp{\left(\frac{-\vE^{b\, 2} + 2\vE^{a}\cdot \vE^{b} -\vE^{b\, 2} }{2\sigma_\text{new}^2}\right)}
\ee
for the vector.  We change to spherical coordinates by $\vE^b = E^b(\sin\vartheta \cos\varphi, \sin\vartheta \sin\varphi, \cos\vartheta)$. Without loss of generality we align $\vE^a$ with the z-axis so that $\Delta\vartheta=\vartheta$ represents the total angular precession over the time interval $(N^a,N^b)$. The actual probability (density times volume element) is then
\be
\mathbf{P}(\vE^b | \vE^a) dV = \frac{1}{(\sqrt{2\pi}\sigma_\text{new})^3} \exp{\left(\frac{-E^{b\, 2} + 2 \cos\Delta\vartheta E^{a} E^{b} -E^{a\, 2} }{2\sigma_\text{new}^2}\right)} \sin\Delta\vartheta E^{b\,2} d\Delta\vartheta d\varphi dE^b.
\label{generalprob}
\ee
To get the probability function for the total precession $\Delta\vartheta$, we integrate over $\varphi\in(0,2\pi)$ and $E^b\in(0,\infty)$, which can be done analytically, and get
\be
\mathbf{P}(\Delta \vartheta | E^a) = \frac{\sin\Delta\vartheta}{2\sqrt{\pi}} \exp\left( -\frac{E^{a\,2}}{2\sigma_\text{new}^2} \right) \left( 2Q + \sqrt{\pi} e^{Q^2}(1+2Q^2)(1+\text{Erf}(Q)) \right),
\label{probability2}
\ee
where we have defined
\be
Q = \frac{E^a \cos\Delta\vartheta}{\sqrt{2}\sigma_\text{new}}
\ee
and the error function
\be
\text{Erf}(Q) = \frac{2}{\sqrt{\pi}} \int_0^Q e^{-x^2} dx.
\ee
The conditional probability is positive definite on the interval ($0,\pi$) where $\Delta\vartheta$ is defined and is properly normalized to unity 
\be
\int_0^\pi \mathbf{P}(\Delta \vartheta | E^a) d\Delta\vartheta =1.
\ee 
This is the probability that $\vE$ will precess an angle $\Delta \vartheta$ during the time ($N^a,N^b$) given that the magnitude is $E^a$ at $N^a$.  However, what we are really after is not the conditional probability given a specific realization at $N^a$, but instead the typical precession during the time interval $(N^a, N^b)$ (without any assumption about the magnitude at $N^a$).  This distribution is given by integrating over all magnitudes $E^a$ weighted by their probability
\be
\mathbf{P}(\Delta\vartheta) = \int_0^\infty \mathbf{P} (E^a) \mathbf{P}(\Delta\vartheta|E^a) dE^a,
\label{probability3}
\ee 
where $\mathbf{P} (E^a)$ is the chi distribution (\ref{chisquared}) (with the substitution $\sigma\rightarrow\sigma_a$ ) and $\mathbf{P}(\Delta\vartheta|E^a)$ is given by (\ref{probability2}). 
This is our sought after probability which characterizes the total precession of $\vE$ between two arbitrary times $N^a$ and $N^b$.  To extract some useful information from it, such as the mean rotation
\be
\left< \Delta\vartheta \right> = \int_0^\pi \Delta\vartheta \mathbf{P}(\Delta\vartheta) d\Delta\vartheta,
\ee
the formidable integral (\ref{probability3}) must be solved numerically. 
\begin{figure}[tbph]
\begin{center}
\includegraphics[width=0.93\textwidth]{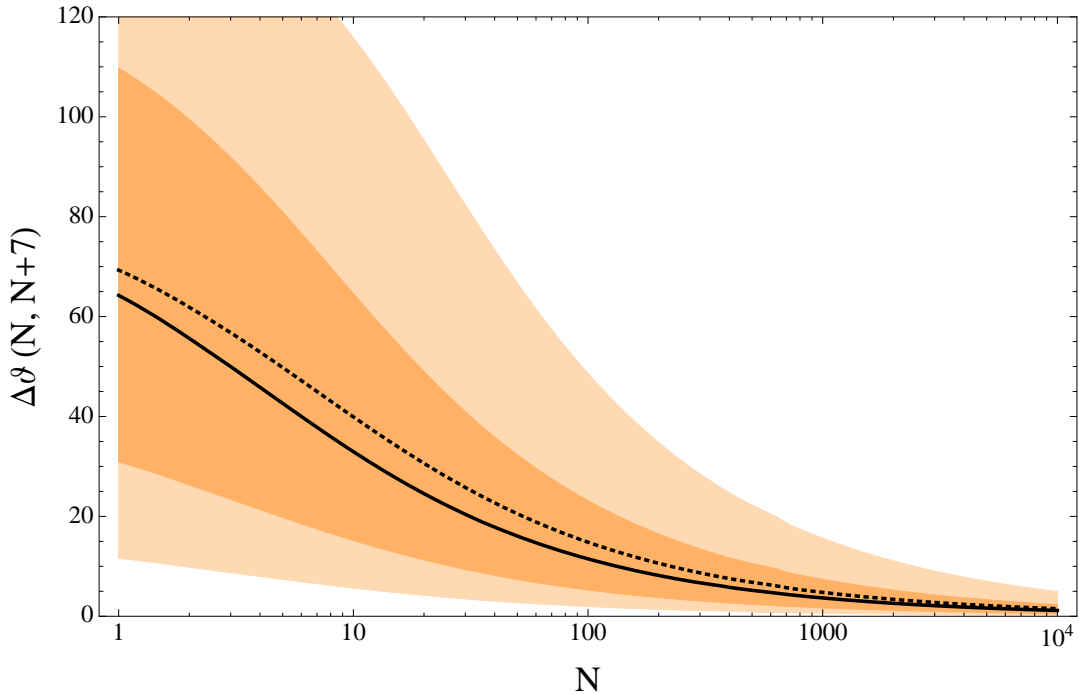}
\end{center}
\caption{The dotted black line is the mean precession $\left<\Delta\vartheta\right>$ over 7 e-folds, while the black continuous line is the same for the median value. The dark and light orange regions represent the $68 \%$ and $95 \%$ confidence intervals, respectively, both centered around the median.}
\label{figstats}
\end{figure}

Figure \ref{figstats} shows the (statistics of the) precession over seven e-folds as a function of time; this roughly corresponds to the time interval over which the CMB modes traverse the horizon ($1$Gpc$/1$Mpc $\simeq e^7$). More specifically, the dotted black line in the upper plot shows the mean value of the precession, $\left< \Delta\vartheta\right>$, over the time interval ($N$, $N\!+\!7$) for $N\in(0,10^4)$, whereas the continuous line is the same for the median value of $\Delta\vartheta$. The dark orange region is the $68\%$ confidence interval, while the $95\%$ confidence interval is in light orange.

We see that the precession of $\vE_{IR}$ is rather significant unless inflation had lasted for an extremely long time before the CMB modes crossed the horizon.  Even if inflation had lasted already $100 [1000]$ e-folds when the CMB modes left the horizon, the mean precession of $\vE_{IR}$ over the next seven e-folds would be $14.8$ [$4.8$] degrees. The relatively weak dependence on $N$ is due to the fact that the mean magnitude scales as $\sqrt{N}$.  Only for $N\gtrsim10^4$ we can be $68\%$ [$95\%$] confident that the precession over one e-fold is in the narrow range (0.5, 2.3) [($0.2, 5.0$)] degrees.  

Next, we derive the PDF for the change of the magnitude $E$ using a similar procedure. Again we start from (\ref{generalprob}), this time integrating over $\varphi\in(0,2\pi)$ and $\vartheta\in(0,\pi)$, to find 
\be
\mathbf{P}(E^b|E^a) = \frac{1}{\sqrt{2\pi}\sigma_\text{new}} \frac{E^b}{E^a} \left( \exp \left[  \frac{2E^aE^b}{\sigma^2_\text{new}} \right] -1\right) \exp \left[ - \frac{(E^a+E^b)^2}{2\sigma_\text{new}^2} \right].
\label{heioghaa}
\ee
We are interested in the distribution of the parameter $X=E^b/E^a$. The conditional probability $\mathbf P(X|E^a)$ is obtained by a simple change of variables in (\ref{heioghaa}). Again, the transition from $\mathbf P(X|E^a)$ to $\mathbf P(X)$  is given by the convolution (\ref{probability3}) which can be done analytically in this case:
\be
\mathbf P(X) = \frac{16}{\pi} \frac{(r^2-1)^{3/2} X^2(X^2+r^2)}{\left[(X^2+r^2)^2-4X^2\right]^2}, 
\label{PX}
\ee
where $r=\sigma_b/\sigma_a>1$ and the normalization is $\int_0^\infty P(X)dX=1$.  We shall employ this result in the next section.

\subsection{CMB/LSS \label{chcmblss}}

Here we shall employ the statistics derived above to give the prior for the expected deviation in primordial parameters. We shall discuss the single vector case and the multivector case separately since the latter requires a Monte-Carlo approach. In both cases, we are primarily interested in the deviation of the amplitude of the quadrupole. We define
\be
\delta = \frac{|g^{(LSS)}|-|g^{(CMB)}|}{|g^{(CMB)}|},
\label{delta}
\ee
which is positive if the detected anisotropy is stronger in LSS than CMB data. The probability distribution of $\delta$ depends on the redshift cut-off in the data sets.  The search for anisotropies in the LSS data carried out in \cite{Pullen:2010} is based on the Sloan Digital Sky Survey with a cut-off in the galaxy catalogue at red-shift $z=0.6$. Euclid, however, will be able to measure galaxies and clusters of galaxies out to redshifts $\sim 2$. For concreteness and simplicity we shall here assume that the comoving CMB horizon is a factor $e$ larger than the comoving LSS horizon.  With a CMB at redshift $1100$ this corresponds to a LSS cut-off at redshift $1.9$ which is within reach of Euclid.

\subsubsection{A single vector \label{single}}
\begin{figure}[tbph]
\begin{center}
\includegraphics[width=0.9\textwidth]{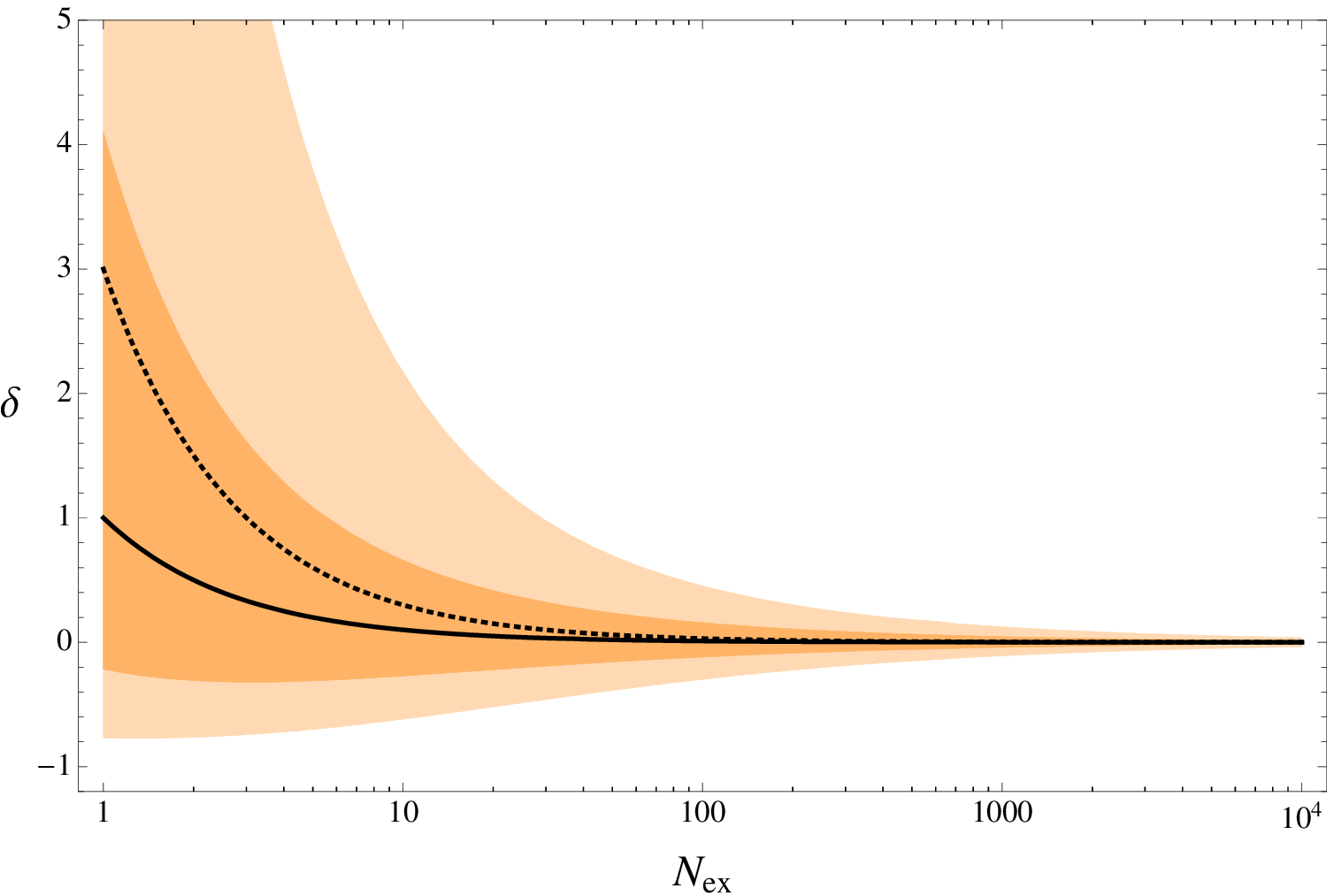}
\includegraphics[width=0.9\textwidth]{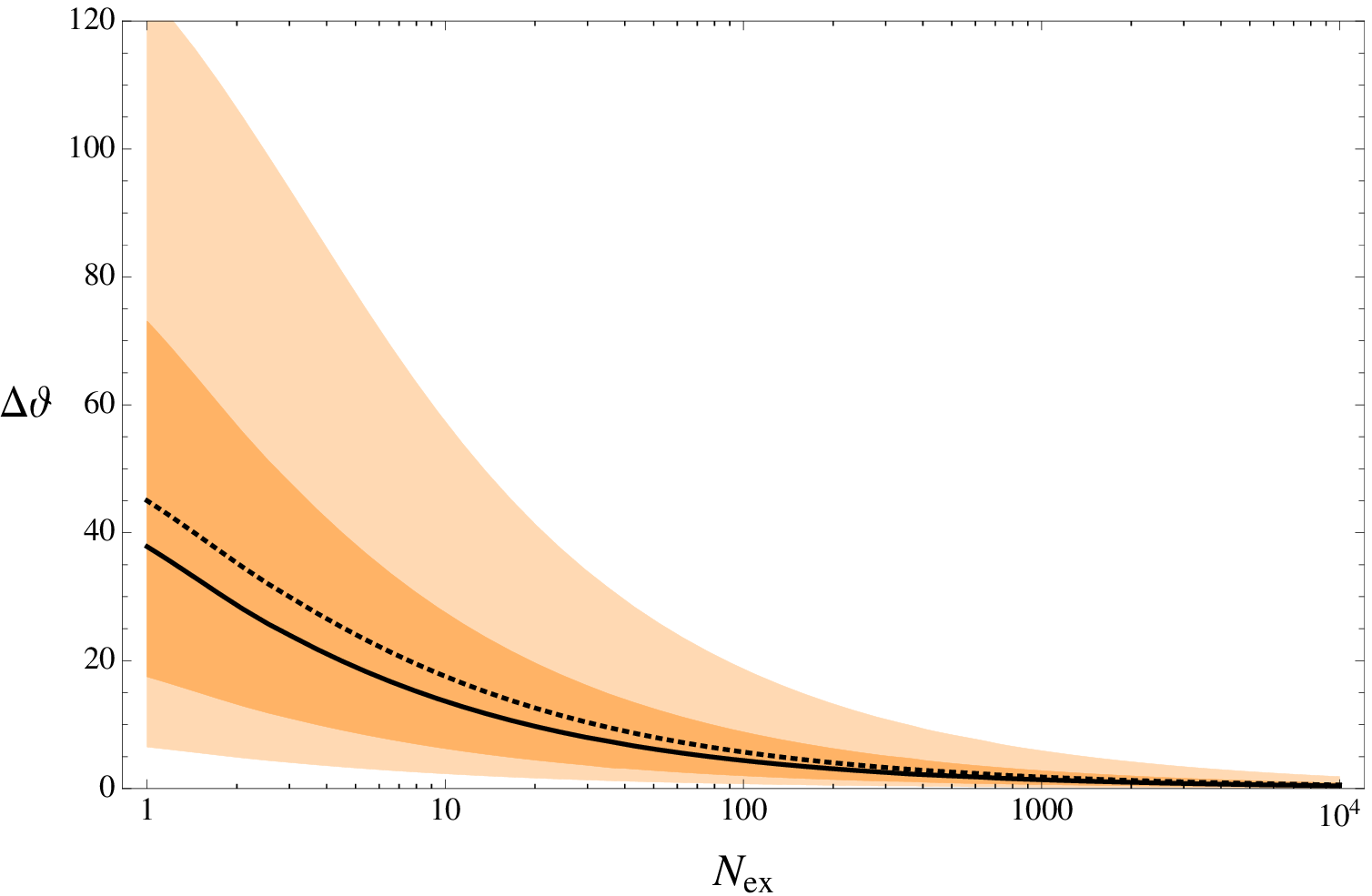}
\end{center}
\caption{Prior for the deviation in amplitude of quadrupole (upper panel) and preferred direction (lower panel) between detections based on CMB and LSS (for single vector models). $\Delta \varphi$ is measured in degrees. The dotted black line is the ensemble mean ($\left<\delta\right>$ and $\left<\Delta\vartheta\right>$ in the upper and lower panel, respectively), while the black continuous line represents the median value. The dark and light orange regions represent the $68 \%$ and $95 \%$ confidence intervals, respectively.}
\label{cmblss2}
\end{figure}

In models with a single gauge vector it is straightforward to employ the statistics derived above to obtain the prior for the deviation in amplitude $g$ and preferred direction $\hat{\mathbf n}$. The former is parametrized by $\delta$ introduced in (\ref{delta}) and the latter by $\Delta\vartheta$ defined
\be
\hat{\mathbf{n}}^{(CMB)} \cdot \hat{\mathbf{n}}^{(LSS)} = \cos\Delta\vartheta. 
\ee
Let us start with  $\delta$ which is determined by the evolution of the magnitude $E$. By a change of variable in (\ref{PX}) we obtain the PDF
\be
\mathbf{P} (\delta) = \frac{8}{\pi} (r^2-1)^{3/2}  \frac{(\delta+r^2+1)\sqrt{\delta+1}}{\left[ (\delta+r^2+1)^2-4(\delta+1) \right]^2},
\ee
where $r^2=(N_\text{ex}+1)/N_\text{ex}$ in agreement with the redshift cut-offs discussed above. This function has the normalization
\be
\int_{-1}^\infty \mathbf{P} (\delta) d\delta = 1.
\ee
In the upper panel of figure \ref{cmblss2} we show the mean, median, $68\%$ and $95\%$ confidence intervals of $\delta$ as function of the free parameter $N_\text{ex}$. For generality we show $N_\text{ex}$ in the large interval $(1,10^4)$ although we know it must be small phenomenologically (section \ref{chplanck}).  As expected it is likely to detect a stronger amplitude in LSS than CMB data, namely the probability for $\delta>0$ is greater than for $\delta<0$. However, it is still possible to measure the strongest $g$ in the CMB as the entire region below the $68\%$ confidence interval corresponds to negative $\delta$. In such realizations gauge modes with comoving wavelength larger than the LSS bubble, but smaller than the CMB bubble, happen to partly cancel out the sum of large scale modes (wavelength larger than the CMB bubble). For $N_\text{ex}=5$, which evades the Planck limit, the median and mean (ensemble average) are 0.20 and 0.60, respectively, while the $68\%$ and $95\%$ confidence intervals are $(-0.31, 1.1)$ and $(-0.70, 3.8)$. The actual deviation measured by an observer depends on his position in Bubbland and corresponds to a single realization drawn from this statistics. In figure \ref{cmblss2} we show the same statistics for the deviation in preferred direction $\Delta\varphi$ measured in degrees. Here we assumed that $\Delta\varphi$ is set by the the precession of $\vE_{IR}$ over a single efold which we calculated numerically using (\ref{probability3}).  For $N_\text{ex}=5$ the median and mean are 19 and 24, respectively, while the $68\%$ and $95\%$ confidence intervals are $(8.7, 38)$ and $(3.2, 78)$.

A comment is in order here.  In this analysis we have ignored the existence of the background solution $\vE_0$ alongside $\vE_{IR}$.  The motivation for doing so comes from the fact that all the new dynamics and interesting observational results come from the stochastic component rather than the background constant component; that is, $\vE_0$ is a truly homogenous vector which does not rotate. The results we depicted so far are mathematically valid in the case the IR component contributes the most, but it is a bit unrealistic that our quantitative conclusions apply exactly in that case; of course qualitatively the picture we describe is still perfectly accurate.  Thus, even though we are, technically speaking, limited to a subdominant $\vE_{IR}$ over $\vE_0$ (corresponding to a small $N_\text{ex}$) in which case $\Delta\varphi$ will be suppressed compared to our estimate, we decided to retain our approximation for its mathematical simplicity and ease of interpretation in terms of the underlying physics.  In this case we are able to track the physical results to the exact dynamical mechanism which triggers them.

\subsubsection{A triad of vectors \label{triad}}
\begin{figure}[tbph]
\begin{center}
\includegraphics[width=1.00\textwidth]{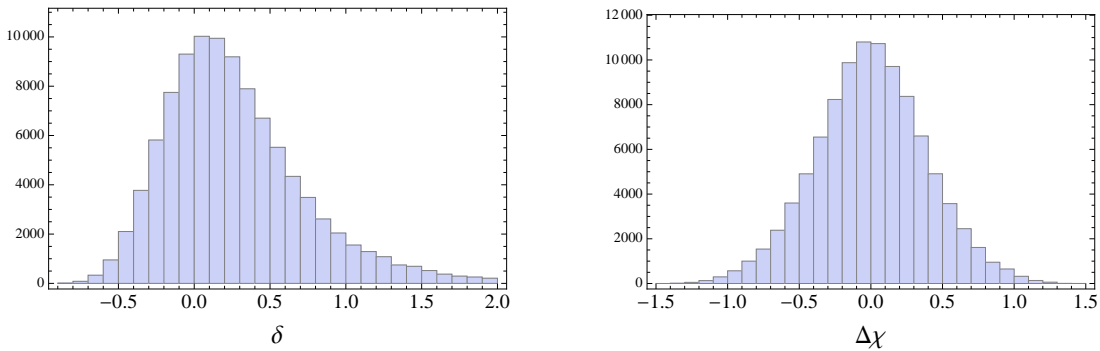}
\end{center}
\caption{Distribution of $\delta$ and $\Delta \chi$ for $10^5$ realizations with $N_\text{ex}=5$.}
\label{figNex5}
\end{figure}

\begin{figure}[tbph]
\begin{center}
\includegraphics[width=1.00\textwidth]{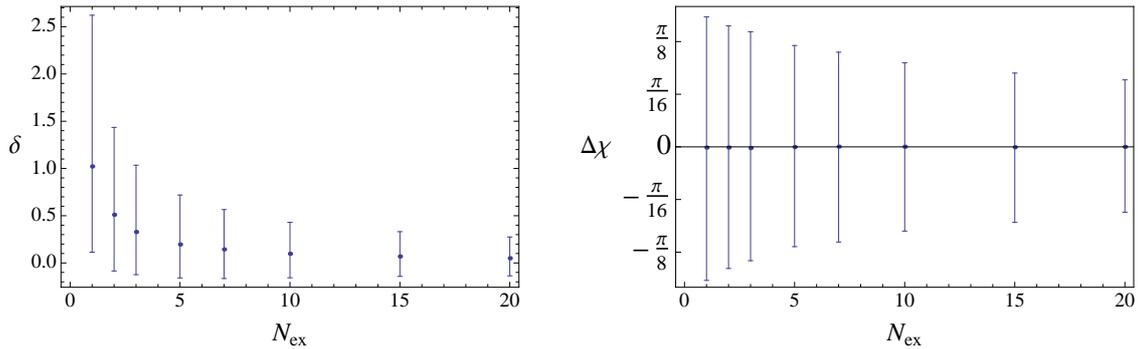}
\end{center}
\caption{Median and $68\%$ error bars for $\delta$ and $\Delta \chi$ as a function of $N_\text{ex}$ ($10^4$ realizations for each value of $N_\text{ex}$). }
\label{figGandDeltaChi}
\end{figure}

As a proof of concept for multi-vector models, we consider here a triad ($n=3$) of vectors.  In addition to $\delta$ defined above we shall also consider the change in the shape parameter
\be
\Delta \chi = \chi^{LSS}-\chi^{CMB}.
\ee
In figure \ref{figNex5} we show the distribution of $\delta$ and $\Delta \chi$ for $10^5$ Monte Carlo realizations with $N_\text{ex}=5$.  As in the single vector case we note that $\delta>0$ is statistically favoured; it is expected to see the largest amplitude in LSS.   In figure \ref{figGandDeltaChi} we show the median value and $68\%$ error bars for a selection of $N_\text{ex}$ on the interval ($1,20$). As expected both parameters decrease with increasing $N_\text{ex}$.\footnote{The imprint in the CMB is dictated by the status of the triad at the time $\tau_0=-1/\mathcal{H}_0$ (roughly), whereas the imprint in the LSS is dictated by the status of the same triad one e-fold later. Then, if $N_\text{ex}$ is large, the relative change over one e-fold will be smaller.} We also note that $\Delta\chi$ is symmetric around $\Delta \chi=0$: the probability for having the largest $\chi$ is democratically shared between the LSS and CMB.  This is also expected since the probability distribution of $\chi$ is independent of $N_\text{ex}$ (see the PDF's reported in \cite{Thorsrud:2013mma}).

\section{Conclusion \label{chconc}}

The Universe generated by inflation might well be much bigger than the patch we observe today.  If this is so, some of the observables we use to describe it may not reflect the true, full-Universe values --- that is, they may be biased. Recently there has been a growing literature, in various contexts, on implications of infrared fluctuations for local correlators, see for instance \cite{Nelson:2012, Nurmi:2013, LoVerde:2013, LoVerde:2013b} for the multi-scalar or isocurvature context. 
 
We have analysed here the statistics of the curvature perturbation $\zeta$ in inflationary models with massless, ghost-free gauge vector kinetic coupling of the type $I(\phi)^2 F^2$.  We specialized to the $\left<I\right> \propto 1/a^2$ case in order to obtain flat gauge spectra --- this solution is in fact a classical attractor for a large class of models for $I(\phi)$.  We have shown how this bias then depends on the local background value of the gauge fields, which include contributions from all modes with wavelength $k \lesssim \cH_0$.  In particular, a quadrupole correction emerges; we have developed a complete parametrization of this correction, that is, we have shown how the most general outcome of the model can be recast in terms of two quantities, an amplitude $g(k)$ and a scale-independent shape $\chi$.

The commonly assumed axial symmetry for the quadrupolar correction ($\chi=0$) is broken by the dynamics of the sum of classicalized gauge modes which make up the infrared vector $\vE_{IR}$.  In fact, the infrared vector experiences a statistical precession phenomenon: it behaves as a random walk in the space of all possible orientations as inflation proceeds, and its imprint is captured by the ampler two-parameter set. For the single-field case we have demonstrated how this correction is in fact close to axisymmetric ($\chi\simeq0$), i.e., one parameter (the amplitude $g$) is usually sufficient to describe the deviation from isotropy.  If there are multiple identical gauge fields instead the symmetry is definitely broken, and the new shape parameter $\chi$ then leads to distinct new signatures.

We have studied the statistics of the quadrupole analytically as well as through Monte Carlo realizations.  The only free quantity in the model is $N_\text{ex}$: the ``extra'' e-folds of inflation in addition to $60$ or so needed in canonical models. For a fixed total duration of inflation it is then possible to determine --- statistically --- what the expected distribution of the quadrupolar parameters would be.  We find that while with a $N_\text{ex}$ of order $1$ the amplitude $g$ stays relatively small, once $N_\text{ex}\gg1$ the distribution rapidly flattens to include appreciable probabilities also for large quadrupoles. Specifically we find that $g_0<0.02$ to $68\%$ ($95\%$) confidence for $N_\text{ex}\simeq 8$ ($N_\text{ex}\simeq 4$).  Another way to look at it is that for $N_\text{ex}\approx17 [76]$, an acceptable 2\% level is realized only less than 32\% [5\%] of the times, that is, the model is $1\sigma [2\sigma]$ away from observations.

Interestingly, the effects of the dynamics of the IR vector background also translates into a prior for the expected deviation between measurents based on different cosmological datasets once such datasets probe different patches.  That is, a cosmological probe going out to a given redshift will in general measure a different primordial quadrupole $(g, \chi)$ than another observable reaching higher $z$. This happens because imprints in local correlators are dictated by the status of the vector background (which is assumed to decay away once inflation ends) at the time $\tau = -1/k_\text{cut}$, where $1/k_\text{cut}$ is the comoving radius of the patch being probed (by an observer today). As we have shown, the vector changes quite appreciably (both in magnitude and direction) over 1 e-fold if $N_\text{ex}$ is within the phenomenological limits discussed above.  In fact a large scale structure survey with redshift cut-off at $z=1.9$ (within reach of Euclid) is sensitive to the status of the IR vector exactly one e-fold later than the $z=1100$ cut-off of the CMB. We found, as expected, that it is more likely to measure a stronger anisotropy in LSS surveys than in the CMB (since the infrared vector had more time to build up stochastically), and that the offset between observed preferred directions can be quite large. To mention an example, for the single vector model with $N_\text{ex}=5$, the median and mean (ensemble average) for the weighted offset of the amplitude, $\delta$, are 0.20 and 0.60, respectively, with $68\%$ and $95\%$ confidence intervals $(-0.31, 1.1)$ and $(-0.70, 3.8)$. 

The results we have obtained in this paper rely on the split $\vE_\text{cl}=\vE_0+\vE_{IR}$ and are mathematically valid in the case the IR component contributes the most; since it is somewaht unrealistic that this is the case, our quantitative conclusions will not apply exactly in nature. Of course, a fully consistent approach requires introducing stochastic noise, both gauge field noise and scalar field noise, into the (classical) field equations and deal with the full solutions for the inflaton as well as the vector (which in general can not be expected to coincide with $\vE_0+\vE_\text{IR}$) --- a task which is far beyond the scope of this paper, and in fact poorly understood even in simpler cases of single field inflation. Nonetheless, we believe the picture we have described in this paper is sufficiently accurate to be relevant --- for sure qualitatively, but hopefully also quantitatively. Especially the multi-vector case is interesting in the context of this issue, since the background attractor is free of anisotropic stress and hence the only anisotropic correction to the power spectrum comes from the infrared component \cite{Thorsrud:2013mma}.  In that case our results are mathematically correct also in the safe regime $|\vE_{IR}|\ll|\vE_0|$.  Furthermore, the parametrization we have developed for the power spectrum is model-independent and from the phenomenological point of view we believe it is important to constrain a general quadrupole $(g,\chi)$ experimentally without the presumption of axial symmetry ($\chi=0$): in particular it would be interesting to see how the limit on the amplitude $g$ depends on the shape parameter $\chi$.

\section*{Acknowledgements}
We would like to thank Jaiseung Kim for helpful correspondence regarding the Planck limit reported in \cite{komatsu}.
FU is supported by IISN project No.\ 4.4502.13 and Belgian Science Policy under IAP VII/37.  He also thanks the University of Oslo for kind hospitality while this work was conceived and completed. DFM is supported by the Research Council of Norway FRINAT grant 197251/V30.

\appendix

\section{Uniqueness of the parameterization \label{chproof}}
In section \ref{chparameterizations} we developed a two dimensional parametrization for power spectrums with a general quadrupole correction.  Here we present the details of the mapping from the five parameters $\{b_{2m}\}$ in (\ref{Qnotrotated}) to the two parameters $\{g,\chi\}$ of (\ref{scaleinvariant}). Our main goal is to prove that the mapping is unique:  for each set $\{b_{2m}\}$ there is one and only one doublet $\{g,\chi\}$.

In this paper we use the following real orthonormal basis for the spherical harmonics (only $l=0$ and $l=2$ are relevant for us):
\begin{align}
Y_{00} &= \frac{1}{2} \sqrt{\frac{1}{\pi}} , \label{Y0} \\
Y_{2m} &= \frac{1}{4} \sqrt{\frac{15}{\pi}} \left\{  \sin{^2\vartheta}  \sin{2\varphi},- \sin{2\vartheta} \sin{\varphi}, \frac{1}{\sqrt{3} } \left(-1+3 \cos{^2\vartheta}\right), -  \sin{2\vartheta}\cos{\varphi}, \sin{^2\vartheta} \cos{2\varphi}  \right\}. \label{Y2}
\end{align}
A general quadrupole is described by the state vector $(b_{2m})$.  An arbitrary rotation of the sphere can be obtained by a rotation $\psi_3$ around the $z$-axis, followed by a rotation $\psi_2$ around the $y$-axis and another rotation $\psi_1$ around the $z$-axis:
\be
\left(\tilde b_{2m}\right) = [R(\psi_1,\psi_2,\psi_3)] (b_{2m}) = \left[R_z(\psi_1)\right] \left[R_y(\psi_2)\right] \left[R_z(\psi_3)\right] \left(b_{2m}\right), 
\ee
where the active, right-handed rotation matrices are defined \cite{matrices}: 
\be
\begin{split}
\left[R_y(\psi)\right] &=
\begin{bmatrix} 
\cos\psi & -\sin\psi & 0 & 0 & 0 \\ 
\sin\psi & \cos\psi & 0& 0& 0 \\
0 & 0 & (1+3\cos 2\psi)/4& (\sqrt{3}\sin2\psi) /2& \sqrt{3}(1-\cos2\psi)/4 \\
0 & 0 & -(\sqrt{3}\sin2\psi)/2 & \cos 2\psi &  (\sin2\psi)/2 \\
0 & 0 & \sqrt{3}(1-\cos2\psi)/4 & -(\sin2\psi)/2 & (3+\cos2\psi)/4 \\
\end{bmatrix}, \\
\left[R_z(\psi)\right] &=
\begin{bmatrix} 
\cos 2\psi & 0 & 0 & 0 & \sin 2\psi \\ 
0 & \cos\psi & 0& \sin\psi& 0 \\
0 & 0 & 1 & 0 & 0 \\
0 & -\sin\psi & 0 & \cos\psi &  0 \\
-\sin2\psi & 0 & 0 & 0 & \cos2\psi \\
\end{bmatrix}.
\end{split}
\ee
In order to write the quadrupole as a linear combination of $Y_{20}$ and $Y_{21}$, we choose the three Euler angles ($\psi_1, \psi_2, \psi_3$) so that $\tilde b_{2m}=0$ for $m=(-2,-1,2)$.

However, the Euler coordinates that satisfy these conditions are not unique.  Suppose that an arbitrary vector $(b_{2m})=(b_{2-2},b_{2-1},b_{20},b_{21},b_{22})$ is rotated to $(\tilde b_{2m})=(0,0,\tilde b_{20}, \tilde b_{21},0)$. We can perform another non-trivial rotation ($\tilde\psi_1, \tilde\psi_2, \tilde\psi_3$) so that the state vector keeps the same form $(\tilde{\tilde{b}}_{2m})=(0,0,\tilde{\tilde{b}}_{20},\tilde{\tilde{b}}_{21},0)$. To develop a unique parametrization we must identify all such rotations ($\tilde\psi_1, \tilde\psi_2, \tilde\psi_3$).  We define $(\alpha_1,\alpha_2,\alpha_3)=(\sin{\tilde \psi_1},\sin{\tilde\psi_2},\sin{\tilde\psi_3})$ and write down the following set of equations from the above matrices:
\begin{align}
0&=\alpha_2\Bigg[ -\tilde b_{21} \alpha_3 \nonumber \\ & \qquad\qquad + \alpha_1 \left(  2\tilde b_{21} \alpha_1 \alpha_3 + \sqrt{1-\alpha_1^2} \left( \sqrt{3}\tilde b_{20} \alpha_2-2\tilde b_{21}\sqrt{1-\alpha_2^2}\sqrt{1-\alpha_3^2}\right)  \right)  \Bigg], \label{sys1} \\
0&= -\tilde b_{20} \alpha_1 \alpha_2 \sqrt{3-3\alpha_2^2}+\tilde b_{21}\alpha_3\sqrt{1-\alpha_1^2}\sqrt{1-\alpha_2^2} + \tilde b_{21} \alpha_1(1-2\alpha_2^2) \sqrt{1-\alpha_3^2}, \label{sys2} \\
0&=\alpha_2 \left[ \frac{\sqrt{3}}{2} \tilde b_{20}\alpha_2(1-2\alpha_1^2) + 2\alpha_1 \alpha_3 \tilde b_{21} \sqrt{1-\alpha_1^2} - \tilde b_{21} (1-2\alpha_1^2) \sqrt{1-\alpha_2^2} \sqrt{1-\alpha_3^2}   \right]. \label{sys3} 
\end{align}
In these equations we used positive roots  $\cos \tilde \psi_i = +\sqrt{1-\alpha_i^2}$ for $i=(1,2,3)$, so that $\tilde \psi_i$ is restricted to the range $[-\pi/2,\pi/2]$. To find the complete set of solutions in the full range $[-\pi,\pi]$ we considered all eight variations of the system (\ref{sys1})-(\ref{sys3}) obtained by the substitutions $\cos \tilde \psi_i = \pm \sqrt{1-\alpha_i^2}$. There are 3 independent non-trivial types of solutions
\begin{align}
(+,-,+):& \quad \alpha_1=\alpha_3,  \quad \alpha_2=0, \label{sol1} \\
(+,+,+):& \quad \alpha_1=\pm1, \quad \alpha_2=0, \quad \alpha_3=\pm1, \label{sol2} \\
(+,+,+):& \quad \alpha_1=0, \quad \alpha_2 = \pm \frac{2\tilde b_{21}}{\sqrt{3|\tilde b_{20}|^2+4|\tilde b_{21}|^2}} \; \text{for} \; \tilde b_{20} \gtrless 0, \quad \alpha_3=0, \label{sol3}
\end{align}
where the parentheses indicate the sign conventions for $\cos \tilde \psi_1$, $\cos \tilde \psi_2$ and $\cos \tilde \psi_3$, respectively. The first solution correspond to a rotation $\arcsin(\alpha_3)\in[-\pi/2,\pi/2,]$ around the $z$ axis, followed by a rotation $\pi$ around the $y$ axis followed by another rotation $\arcsin(\alpha_3)$ around the $z$ axis. The first and third rotation cancels so that the solution (\ref{sol1}) correspond to a rotation $\pi$ around the $y$ axis. Solution (\ref{sol2}) describes a rotation $\pm \pi/2$ around the $\tilde z$-axis followed by another rotation $\pm \pi/2$ around the same axis, adding up to a total rotation $\pi$ around the $\tilde z$-axis. The third solution (\ref{sol3}) represents a rotation around the $\tilde y$-axis with an angle $\in [-\pi/2,\pi/2]$ that depends on $\tilde b_{20}$ and $\tilde b_{21}$. 

After an arbitrary rotation consistent with the sign convention $(+,+,+)$ the new state vector has the following non-vanishing components:
\begin{align}
\tilde{\tilde b}_{20} &=  \tilde b_{20}\left(1 -\frac{3}{2} \alpha_2^2 \right) + \tilde b_{21} \sqrt{3} \alpha_2 \sqrt{1-\alpha_2^2} \sqrt{1-\alpha_3^2}, \label{ola1} \\
\tilde{\tilde b}_{21} &= -\tilde b_{20} \sqrt{3} \sqrt{1-\alpha_1^2} \alpha_2 \sqrt{1-\alpha_2^2} + \tilde b_{21} \sqrt{1-\alpha_1^2}(1-2\alpha_2^2)\sqrt{1-\alpha_3^2} -\tilde b_{21} \alpha_1 \sqrt{1-\alpha_2^2} \alpha_3. \label{ola2}
\end{align}
For the sign convention $(+,-,+)$ the corresponding equations are obtained by substituting $\sqrt{1-\alpha_2^2} \rightarrow -\sqrt{1-\alpha_2^2}$ in (\ref{ola1})-(\ref{ola2}). Inserting the solution (\ref{sol1}) we see that the state vector is invariant
\be
(\tilde{\tilde b}_{2m}) = (0, 0, \tilde b_{20}, \tilde b_{21}, 0),
\ee 
while if we insert (\ref{sol2}) or  (\ref{sol3}) the new state vector is
\be
(\tilde{\tilde b}_{2m}) = (0, 0, \tilde b_{20}, -\tilde b_{21}, 0).
\ee 
Thus we have showed that the ambiguity in the Euler coordinates that set $\tilde b_{2-2}=\tilde b_{2-1}=\tilde b_{22}=0$ is very restrictive. We have freedom to choose the signature of $\tilde b_{21}$ whereas $\tilde b_{20}$ and $|\tilde b_{21}|$ are unique. To remove the ambiguity we choose 
\be
\text{sign}(\tilde b_{21})=-\text{sign}(\tilde b_{20}).
\label{convention}
\ee
With this convention the mapping from five arbitrary coefficients $\{ b_{2m} \}$ to $\{\tilde b_{20},\tilde b_{21}\}$ is unique: for each set $\{ b_{2m} \}$ there is one and only one doublet $\{\tilde b_{20},\tilde b_{21}\}$.  Using these results the mapping from (\ref{Qnotrotated}) to (\ref{scaleinvariant}) is straight forward:
\begin{align}
\mathcal{P}_\zeta (\mathbf k) &=  P(k) \left( 1 + \sum_{m=-2}^{2} b_{2m} Y_{2m} (\vartheta,\varphi) \right)  \notag \\
&=  P(k) \left( 1 + \tilde b_{20}Y_{20}(\tilde\vartheta,\tilde\varphi) + \tilde b_{21}Y_{21}(\tilde\vartheta,\tilde\varphi) \right)  \notag \\
&= P(k) \left( 1+\frac{3}{4} \sqrt{\frac{5}{\pi}} \left[ \tilde b_{20} (\cos^2{\tilde\vartheta} -1/3) + \sqrt{\frac{1}{3}} (-\tilde b_{21}) \sin{2\tilde\vartheta} \cos{\tilde\varphi}  \right]
   \right)  \notag \\
&= P(k) \left( 1+ \text{sign}(\tilde b_{20}) \frac{3}{4} \sqrt{\frac{5}{\pi}} \left[ |\tilde b_{20}| (\cos^2{\tilde\vartheta} -1/3) + \sqrt{\frac{1}{3}} |\tilde b_{21}| \sin{2\tilde\vartheta} \cos{\tilde\varphi}  \right]
   \right),   \notag
\end{align}
where in the last step we have used (\ref{convention}). Comparing with (\ref{scaleinvariant}), we obtain the definitions of $g$ and $\chi$, see equations (\ref{gdef})-(\ref{chi2def}).  


Finally we remark that although the convention (\ref{convention}) ensures the uniqueness of the mapping $\{ b_{lm} \} \rightarrow \{ \tilde b_{20}, \tilde b_{21} \}$,  it does not fix the coordinate system uniquely. In fact,  there are exactly four coordinate systems corresponding to the map. To see this, note that there are 2 non-trivial operations which leaves the state vector $(\tilde b_{2m})$ invariant. The first one is given by solution (\ref{sol1}). The second one is given by solution (\ref{sol2}) followed by (\ref{sol3}) (or in the opposite direction since they commute). Together with the unit element these operations form a group which can be used to switch between the four coordinate systems.

\bibliographystyle{JHEP}
\bibliography{refs}

\end{document}